\documentclass{pinchcr}   

\usepackage{amsmath}
\usepackage{amsfonts}

\newcommand{\bra}[1]{\left\langle #1\right|}
\newcommand{\ket}[1]{\left|#1\right\rangle}

\newcommand{\mean}[1]{\left\langle #1 \right\rangle}

\newcommand{\ImaginaryPart}{{\rm Im}}

\newcommand{\rp}{\right)}
\newcommand{\lp}{\left(}

\newcommand{\rsb}{\right]}
\newcommand{\lsb}{\left[}
\newcommand{\lbv}{\left|}
\newcommand{\rbv}{\right|}
\newcommand{\lvb}{\lbv}
\newcommand{\rvb}{\rbv}
\newcommand{\comm}[2]{\lsb #1,#2\rsb}

\newcommand{\pref}[1]{(\ref{#1})}
\renewcommand{\eqref}[1]{Eq.~\pref{#1}}
\newcommand{\hc}{\mathrm{h.c.}}

\newcommand{\superop}[1]{{\mathcal #1}}
\newcommand{\sD}{{\superop{D}}}
\newcommand{\sL}{{\superop{L}}}

\newcommand{\tr}[1]{\mathbf{#1}}

\newcommand{\tD}{{\tr{D}}}

\newcommand{\tP}{{\tr{P}}}

\newcommand{\projector}{\Pi}
\newcommand{\proj}[1]{\projector_{#1}}
\newcommand{\dproj}[1]{\dot{\projector}_{#1}}

\newcommand{\Kp}{K'}

\newcommand{\ad}{{a^\dag}}
\newcommand{\ada}{a^\dag a}

\title{Nonlinear oscillators}

 \author{P.~Bertet}
 \affiliation{Quantronics group, Service de Physique de l'\'Etat Condens\'e
(CNRS URA 2464), IRAMIS, DSM, CEA-Saclay, 91191 Gif-sur-Yvette, France }
\author{F.~R.~Ong}
 \affiliation{Quantronics group, Service de Physique de l'\'Etat Condens\'e
(CNRS URA 2464), IRAMIS, DSM, CEA-Saclay, 91191 Gif-sur-Yvette, France }
 \author{M.~Boissonneault}
\affiliation{D\'epartement de Physique, Universit\'e de Sherbrooke, Sherbrooke, Qu\'ebec, Canada, J1K 2R1}
\author{A.~Bolduc}
\affiliation{D\'epartement de Physique, Universit\'e de Sherbrooke, Sherbrooke, Qu\'ebec, Canada, J1K 2R1}
\author{F.~Mallet}
 \affiliation{Quantronics group, Service de Physique de l'\'Etat Condens\'e
(CNRS URA 2464), IRAMIS, DSM, CEA-Saclay, 91191 Gif-sur-Yvette, France }
\author{A.~C.~Doherty}
 \affiliation{School of Physics, The University of Sydney, Sydney, NSW 2006, Australia}
\author{A.~Blais}
\affiliation{D\'epartement de Physique, Universit\'e de Sherbrooke, Sherbrooke, Qu\'ebec, Canada, J1K 2R1}
\author{D.~Vion}
 \affiliation{Quantronics group, Service de Physique de l'\'Etat Condens\'e
(CNRS URA 2464), IRAMIS, DSM, CEA-Saclay, 91191 Gif-sur-Yvette, France }
\author{D.~Esteve}
\affiliation{Quantronics group, Service de Physique de l'\'Etat Condens\'e
(CNRS URA 2464), IRAMIS, DSM, CEA-Saclay, 91191 Gif-sur-Yvette, France }

\authors{9}

\begin{document}

\maintext


\chapter{Circuit quantum electrodynamics with a nonlinear resonator} 
\label{cha:circuit_quantum_electrodynamics_with_a_nonlinear_resonator}
One of the most studied model systems in quantum optics is a two-level atom strongly coupled to a single mode of the electromagnetic field stored in a cavity, a research field named cavity quantum electrodynamics or CQED~\shortcite{haroche_exploring_2006}. This extremely simple quantum system has nevertheless nontrivial quantum dynamics described by the Jaynes-Cummings Hamiltonian~\shortcite{Jaynes1963}. In its implementation at microwave frequencies, it has allowed the observation of many basic concepts of quantum mechanics such as the quantum jumps of the electromagnetic field~\shortcite{guerlin_progressive_2007}, as well as the generation and tomography of nonclassical states~\shortcite{deleglise_reconstruction_2008}. In the context of quantum information, elementary quantum gates have been realized using the cavity as a catalyst for entanglement between atoms passing successively through it~\shortcite{rauschenbeutel_coherent_1999}. CQED has recently received renewed attention due to its implementation with superconducting artificial atoms and coplanar resonators in the so-called circuit quantum electrodynamics (cQED) architecture~\shortcite{Blais2004,Wallraff2004}. In cQED, the couplings can be much stronger than in CQED due to the design flexibility of superconducting circuits and to the enhanced field confinement in one-dimensional cavities, compensating the shorter coherence times of superconducting qubits. This enabled the realization of fundamental quantum physics~\shortcite{hofheinz_synthesizing_2009,ansmann_violation_2009,palacios-laloy_experimental_2010} and quantum information processing~\shortcite{DiCarlo2009} experiments with a degree of control comparable to that obtained in CQED. Even though the physical implementation is different, the system is described by the same Hamiltonian in cQED as in CQED. 

The purpose of this chapter is to investigate the situation where the resonator to which the atom is coupled is made nonlinear with a Kerr-type nonlinearity, causing its energy levels to be nonequidistant. The system is then described by a nonlinear Jaynes-Cummings Hamiltonian. This considerably enriches the physics since a pumped nonlinear resonator displays bistability~\shortcite{Siddiqi2005}, parametric amplification~\shortcite{castellanos-beltran_widely_2007}, and squeezing~\shortcite{castellanos-beltran_amplification_2008}. The interplay of strong coupling and these nonlinear effects constitutes a novel model system for quantum optics that can be implemented experimentally with superconducting circuits. 

This chapter is organized as follows. In a first section we present the system consisting of a superconducting Kerr nonlinear resonator strongly coupled to a transmon qubit~\shortcite{Koch2007}. In the second section, we describe the response of the sole nonlinear resonator to an external drive, with a particular emphasis on the bistable regime. Our main interest is the probability that the resonator switches from its initial dynamical state to a state of higher amplitude under driving by a microwave pulse. We present here both measurements and numerical simulations of this process and obtain quantitative agreement. In the third section, building on our understanding of the switching process, we show how the resonator bistability can be used to perform a high-fidelity readout of the transmon qubit, a crucial task for quantum information processing~\shortcite{mallet_single-shot_2009}. In the course of this readout, the nonlinear resonator is energized by a small microwave field; it is then natural to investigate what is the quantum backaction exerted by the intracavity field on the qubit~\shortcite{Ong2011}. This is the subject of the fourth section.

\section{Presentation of the system} 
\label{sec:presentation_of_the_system}

\subsection{Nonlinear resonator} 
\label{sub:nonlinear_resonator}

\begin{figure}[t]
\begin{center}
   \includegraphics[width=8.35cm,angle=0]{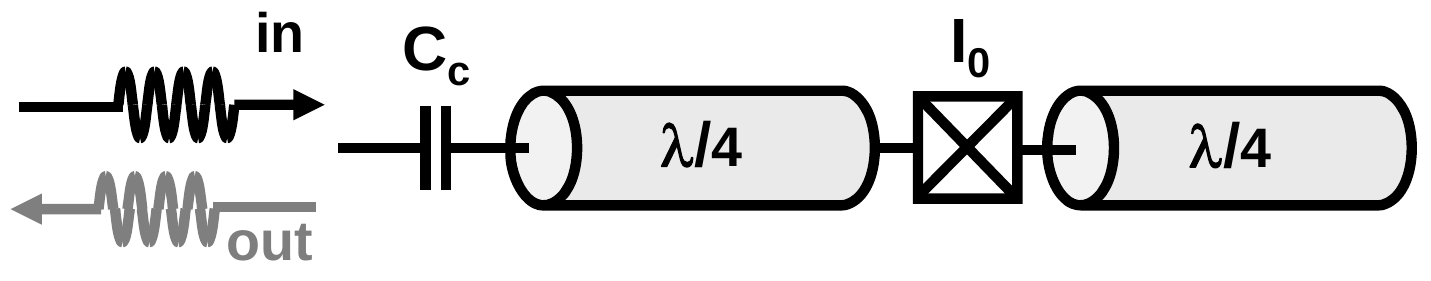}
\end{center}
\caption{Scheme of the nonlinear resonator. A Josephson junction of critical current $I_0$ is embedded in the middle of a $\lambda/2$ resonator. It is coupled to a $50\Omega$ transmission line through a coupling capacitor $C_c$ and probed in reflection by a microwave field.}
\label{figKNR}
\end{figure}

The heart of our device is a superconducting coplanar waveguide resonator~\shortcite{frunzio_fabrication_2005,goppl_coplanar_2008} made nonlinear with a Josephson junction. Linear coplanar waveguide resonators consist of a section of length $L$ of a coplanar waveguide transmission line, supporting modes at frequencies $\tilde{\omega}_n=(n+1) \pi \bar{c} / L$, $\bar{c}$ being the speed of light in the transmission line. In the limit where the quality factor of each mode is high, one can consider only the dynamics of the mode closest in frequency to the system of interest (namely the qubit to which it is coupled), in our case the fundamental mode $n=0$. The resonator Hamiltonian is then $H=\hbar \omega_r a^\dagger a$, where $\omega_r=\tilde{\omega}_0$ and $a$ ($a^\dagger$) is the usual annihilation (creation) operator.

In our experiment a Josephson junction of critical current $I_0$ is inserted in the middle of the transmission line resonator (see Fig.~\ref{figKNR}). A Josephson junction can be considered from an electrical engineering point of view as a lumped dissipationless nonlinear inductance. As a result, the mode structure of the resonator is deeply modified. First, the frequencies of all the resonator modes are differently shifted by the introduction of the lumped inductance, so that they don't follow the simple relation $\tilde{\omega}_n=(n+1) \pi \bar{c} / L$ anymore. Second, the nonlinearity of the Josephson inductance results in nonlinear dynamics for each resonator mode and couplings of the resonator modes with each other. This could make the dynamics of this distributed nonlinear system hardly tractable. However, in the limit where the frequencies of the signals driving the resonator do not generate harmonics that are resonant with higher modes~\shortcite{zakka-bajjani_quantum_2011}, one can approximate the resonator to only one nonlinear mode. This is valid provided the frequency of higher modes is sufficiently different from $(n+1) \pi \bar{c} / L$ (which requires a relatively large lumped element inductance) and provided the frequency of the signals stays close to $\omega_r$ (which requires that the quality factor of the fundamental mode is sufficiently large). All the results presented in the following are obtained within this approximation. More details on the modelling of the distributed nonlinear resonator can be found in~\shortcite{wallquist_selective_2006}.

\begin{figure}[t]
\begin{center}
\includegraphics[width=8.35cm,angle=0]{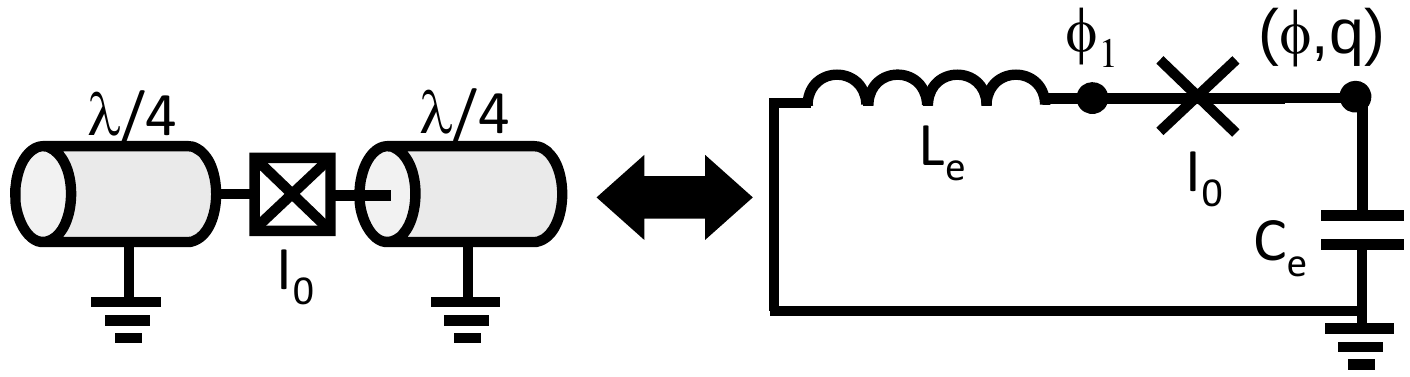}
\end{center}
\caption{Equivalence between the distributed nonlinear resonator and a series combination of an equivalent inductance $L_{\rm e}$, capacitance $C_{\rm e}$ and Josephson junction of critical current $I_0$.
}
\label{figSM}
\end{figure}

We now derive the Hamiltonian of the nonlinear resonator restricted to the fundamental mode, and find explicit expressions for the nonlinear constants~\shortcite{palacios-laloy_tunable_2008,Ong2011}. The first step is to map the distributed nonlinear resonator onto an equivalent series combination of a lumped element inductance $L_{\rm e}$, capacitance $C_{\rm e}$ (which includes the junction capacitance) and Josephson junction of critical current $I_0$, as shown in Fig.~\ref{figSM}~\shortcite{manucharyan_microwave_2007}. In the limit where the Josephson inductance $L_{\rm J}$ is completely negligible compared to the resonator inductance, one can show that $L_{\rm e} = \pi Z_0 / 2 \omega_1$ and $C_{\rm e} = 2 / \pi Z_0 \omega_1$, where $Z_0 \sim $~50 $\Omega$ is the resonator characteristic impedance and $\omega_1$ the resonance frequency in absence of the junction. When $L_{\rm J}$ is not negligible, one has to adjust numerically $L_{\rm e}$ and $C_{\rm e}$ so that the impedance of the equivalent circuit fits the impedance of the distributed resonator. This is the approach that will be used in the following.

Using the notation introduced in Fig.~\ref{figSM}, we obtain the equivalent circuit Hamiltonian
\begin{equation}
	H=\frac{\phi_{1}^{2}}{2L_{\rm e}}-E_{\rm J}\cos\left(\frac{\phi-\phi_{1}}{\varphi_{0}}\right)+\frac{q^{2}}{2C_{\rm e}},
\end{equation}
where $\varphi_{0}=\hbar/2e$ is the reduced superconducting flux quantum. Since the current $I$ flowing through the inductance is the same as that in the junction, we also have
\begin{equation}
	I=\frac{\phi_{1}}{L_{\rm e}}=I_{0}\sin\left(\frac{\phi-\phi_{1}}{\varphi_{0}}\right),
\end{equation}
yielding an implicit relation $\phi_{1}=g(\phi)$ between the two phases. Eliminating $\phi_1$ and expanding $g(\phi)$ in powers of $\phi$, we obtain the nonlinear resonator Hamiltonian to any order of Josephson junction nonlinearity. For instance, to sixth order we find
\begin{equation}
	H=\frac{\phi^{2}}{2L_{\rm t}}+\frac{q^{2}}{2C_{\rm e}}-\frac{1}{24}p^{3}\frac{\phi^{4}}{L_{\rm t}\varphi_{0}^{2}}+\frac{1}{720}p^5 (-9+10p) \frac{\phi^{6}}{L_{\rm t}\varphi_{0}^{4}},
\end{equation}
where $L_{\rm t}=L_{\rm J} + L_{\rm e}$ is the total inductance and $p=L_{\rm J} / L_{\rm t}$ the participation ratio of the Josephson inductance in the total resonator inductance.

This Hamiltonian can be writen in terms of $a^\dag$ and $a$ with $\phi=i \sqrt{\hbar Z_{\rm e}/2} (a-a^{\dagger})$ and $q =\sqrt{\hbar /2 Z_{\rm e}} (a+a^{\dagger})$ with  $Z_{\rm e}=\sqrt{L_{\rm t} / C_{\rm e}}$.  Once expanded, the nonlinear terms $\phi^{4}$ and $\phi^{6}$ yield products of creation and annihilation operators to various powers. Using the rotating-wave approximation, we keep only those with equal number of annihilation and creation operators yielding the Hamiltonian of a Kerr nonlinear resonator (KNR)

\begin{equation}
	\label{eqn:NLO_hamiltonian}
	H_{NL} / \hbar = \omega_r \ada + \frac{K}{2} \ad^2 a^2 + \frac{\Kp}{3}\ad^3 a^3,
\end{equation}

where

\begin{subequations}
\begin{align}
	K &=-\frac{\pi p^{3} \omega_r Z_{\rm e}}{R_{\rm K}}, \\
	K' &= \frac{2}{3p}\frac{K^2}{\omega_r}\left(10p-9\right).
	\end{align}
\end{subequations}

\noindent are the nonlinear constants respectively to third (Kerr constant) and fifth order, $\omega_r = 1/\sqrt{L_{\rm t} C_{\rm e}}$ and $R_K = h/e^2$. Here and below, we have dropped the resulting small correction to $\omega_r$. Note that we will only consider $K' \neq 0$ in the last section of this chapter where it should be taken into account for a quantitative agreement between theory and the data.

In order to allow measurement of the resonator without perturbing its dynamics, the resonator is connected to a $50\,\Omega$ measurement line through a small coupling capacitor $C_c$ that determines its quality factor $Q$, or damping rate $\kappa = \omega_r / Q$. It can be probed by a microwave signal of frequency $\omega_p$ and amplitude $\epsilon_p$, which is reflected at the resonator input [see Figs.~\ref{figKNR} and \ref{figscurves}~a)]. The amplitude and phase of the reflected signal is measured by homodyne detection at room temperature after being amplified by a cryogenic amplifier. More generally, driving of the resonator is described by the Hamiltonian term
\begin{equation}
	\label{eqn:drive_hamiltonian}
	H_d/\hbar = \sum_{d\in\{s,p\}} \epsilon_d e^{-i\omega_d t}\ad + \hc,
\end{equation}
where two different drives have been considered: the pump drive (amplitude $\epsilon_p$ and frequency $\omega_p$) which will describe driving of the resonator close of its resonance frequency and the qubit drive (amplitude $\epsilon_s$ and frequency $\omega_s$) will describe driving of the qubit close to its resonance frequency through the resonator.

The complete resonator dynamics is described in the usual Markov and rotating wave approximations, and for sufficiently weak nonlinearity $K,K'\ll \omega_r$, by the master equation~\shortcite{walls_quantum_2008}
\begin{equation}
	\label{eqn:NLO_masterequation}
	\dot\rho = \sL_r\rho = \frac{-i}{\hbar}\comm{H_{NL}+H_d}{\rho} + \kappa(n_{\rm th}+1)\sD[a]\rho + \kappa n_{\rm th}\sD[\ad]\rho,
\end{equation}
with the dissipator
\begin{equation}
	\label{eqn:dissipator}
	\sD[A]\rho \equiv \frac12 (2 A\rho A^\dag - A^\dag A \rho - \rho A^\dag A),
\end{equation}
and where $n_{\rm th}=1/[\exp(\hbar\omega_r/k_BT)-1]$ is the number of thermal photons in the bath coupled to the resonator at temperature $T$, and $k_B$ is the Boltzmann constant. 


\subsection{Qubit} 
\label{sub:qubit}

\begin{figure}[t]
	\begin{center}
		\includegraphics[width=11cm,angle=0]{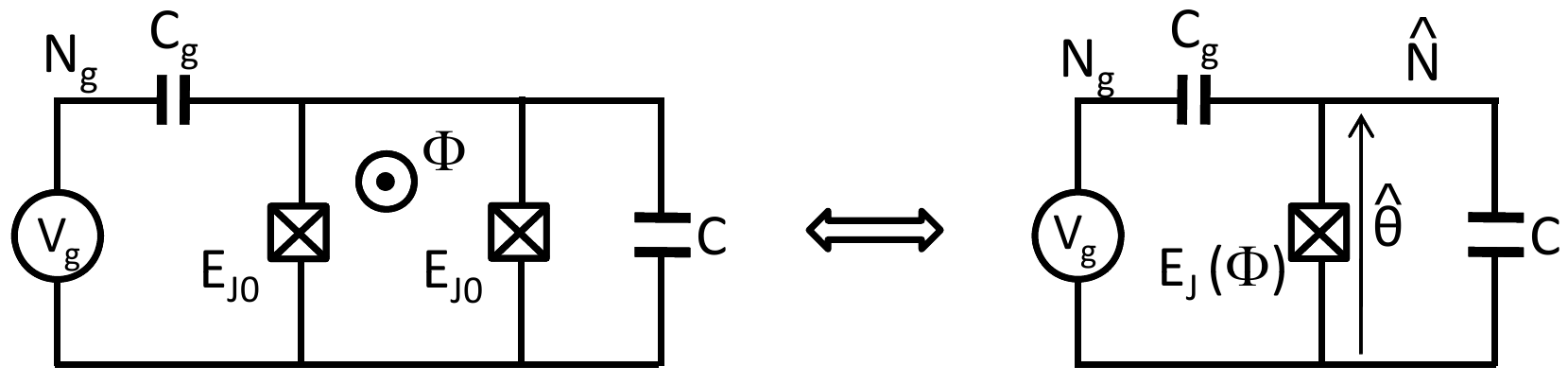}
	\end{center}
	\caption{Circuit model of a split Cooper-pair box made of a superconducting loop including two Josephson junctions of Josephson energy $E_{J0}$ threaded by a flux $\Phi$, shunted by a capacitor $C$, and charge-biased by a voltage source $V_g$ through a gate capacitor $C_g$ yielding a charge bias $N_g=C_g V_g / (2e)$.
}
	\label{figTransmon}
\end{figure}

The artificial atom used in our experiment is a split Cooper-pair box operated in the phase regime (i.e. a transmon qubit, see ~\shortcite{Koch2007,schreier_suppressing_2008,Houck2009}). As illustrated in Fig.~\ref{figTransmon}, it consists of two superconducting islands connected by a superconducting loop (a SQUID) made of two Josephson junctions of Josephson energy $E_{J0}$ and shunted by a capacitor $C$. The SQUID is threaded by a flux $\Phi$ and charge-biased by a voltage source $V_g$ through a gate capacitor $C_g$. Its Hamiltonian is $H=E_{C}\left(N-N_{g}\right)^{2}-E_{J}(\Phi)\cos\left(\theta\right)$, $N$ being the number of Cooper pairs transferred from one island to the other, $N_g=C_g V_g/(2e)$ the charge bias, $\theta$ the gauge-invariant phase difference between the superconductors, $E_C=(2e)^2/2C$ the box charging energy and $E_{J}(\Phi) = 2 E_{J0} \left| \cos (\pi \Phi / \Phi_0) \right|$ the SQUID flux-dependent effective Josephson energy. This Hamiltonian can be diagonalized yielding the qubit eigenstates $\{\ket{i}\}$ with frequencies $\{\omega_i\}$. In this eigenbasis, the Cooper-pair box Hamiltonian restricted to its $M$ first eigenstates is
\begin{equation}
	\label{eqn:mls_hamiltonian}
	H_q/\hbar = \sum_{i=0}^{M-1} \omega_i\proj{i,i} \equiv \proj{\omega}.
\end{equation}
where $\proj{i,j} = \ket{i}\bra{j}$. In the following, we will note $\omega_{i,i+1}=\omega_{i+1}-\omega_i$.

A transmon qubit is a Cooper-pair box operated in the limit where $E_{J}(\Phi) \gg E_{C}$. This has important consequences that are discussed extensively in~\shortcite{Koch2007}. The transmon is essentially insensitive to charge noise, which increases its coherence time considerably. Its eigenfrequencies $\omega_i$ obey approximate simple relations $\hbar \omega_{01} \sim \sqrt{2 E_{J}(\Phi) E_{C}}$ and $\omega_{i+1,i+2} \sim \omega_{i,i+1} - E_{C}/(4 \hbar)$. Its wavefunctions are very similar to those of a harmonic oscillator. This implies interesting properties for the transmon matrix elements: approximately, we have $\bra{i} N \ket{j} \approx \delta_{j,i + 1} \sqrt{i+1} \bra{0} N \ket{1}$ (assuming $j>i$) and similarly $\bra{i} \theta \ket{j} \approx \delta_{j,i + 1} \sqrt{i+1} \bra{0} \theta \ket{1}$.\footnote{These relations are approximate. In the numerical simulations presented in the chapter, the exact energies and couplings were computed by diagonalizing the Cooper-pair box Hamiltonian.}

Since spontaneous decay of the excited states of the transmon occurs mainly by linear coupling of the charge or phase degrees of freedom to the various baths (either electromagnetic or microscopic), this in turn implies that relaxation of the transmon occurs mostly from level $i+1$ to level $i$ (and not to lower-lying levels) with a rate $\gamma_{i+1,i}$ that verifies $\gamma_{i+1,i} \approx (i+1) \gamma$ where $\gamma$ is the spontaneous emission rate of level $\ket{1}$ towards $\ket{0}$. Other uncontrolled processes include low-frequency noise (flux or charge noise mostly) which causes dephasing of level $\ket{i}$ at a rate $\gamma_{\phi,i}$. All these processes are described in the following master equation\footnote{Note that this description for dephasing is insufficient to describe dephasing of the levels $i>1$. A quantitative model of dephasing for these higher levels should take into account the spectrum of the bath causing dephasing. This description is however sufficient for our purpose in this chapter.}
\begin{equation}
	\label{eqn:mls_masterequation}
	\dot\rho = \sL_q \rho = -\frac{i}{\hbar}\comm{H_q}{\rho} + \sum_{i=0}^{M-2}\gamma_{i+1,i} \sD[\proj{i,i+1}]\rho + 2\sum_{i=0}^{M-2}\gamma_{\phi,i} \sD[\proj{i,i}]\rho,
\end{equation}
where we considered the qubit to be at zero temperature. In addition, the qubit state can be manipulated by driving through the resonator [see \eqref{eqn:drive_hamiltonian}].


\subsection{Coupling} 
\label{sub:coupling}

\begin{figure}[t]
\begin{center}
\includegraphics[width=11cm,angle=0]{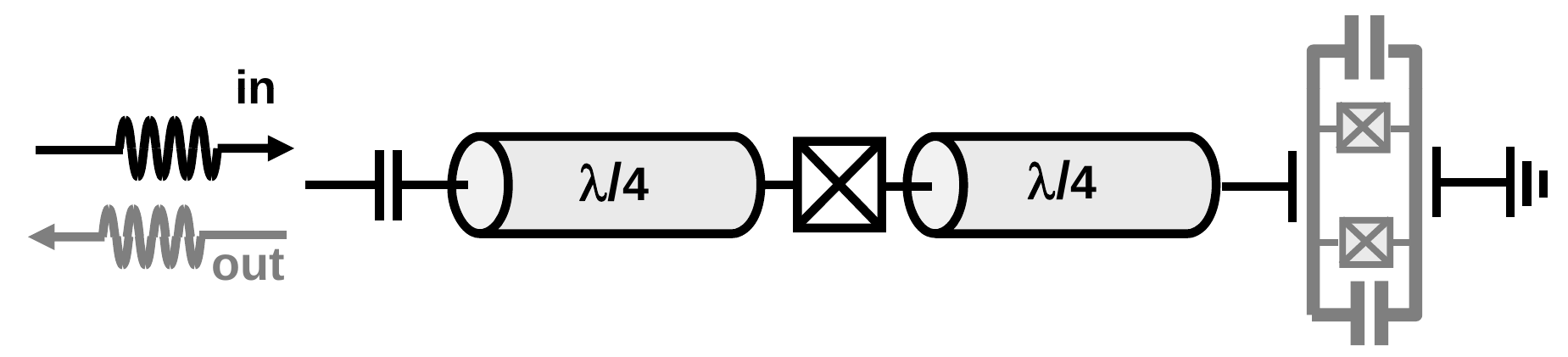}
\end{center}
\caption{Coupled qubit-nonlinear resonator system. The transmon qubit is capacitively coupled to one end of the nonlinear resonator, while the other end is coupled to a $50 \Omega$ measurement line through which microwave pulses of amplitude $\epsilon_p$ (resp. $\epsilon_s$) and frequency $\omega_p$ (resp. $\omega_s$) drive the resonator (resp. the qubit).
}
\label{figqubitKNR}
\end{figure}
In our setup the transmon qubit is capacitively coupled to the resonator by the gate capacitor $C_g$ and located at one end of the resonator as depicted in Fig.~\ref{figqubitKNR}. It is straightforward to show~\shortcite{Koch2007} that such a capacitive transmon-resonator interaction is described by a multilevel Jaynes-Cummings model
\begin{equation}
	H_I/\hbar = \sum_{i<j}^{M-2} g_{i,j} (\ad\proj{i,j} + a\proj{j,i}),
\end{equation}
with $g_{i,j} = 2 e \beta \bra{i} \hat{N} \ket{j} \delta V_0$, $\delta V_0 = \omega_r \sqrt{\hbar Z_0/\pi}$ the vacuum fluctuations of the voltage between the resonator center conductor and ground plane at the qubit location, and $\beta=C_g/(C+C_g)$. Note that we have neglected counter-rotating terms because they oscillate rapidly and only contribute to a slight energy shift. Due to the properties of the transmon matrix elements discussed above $g_{i,j} \approx \delta _{j,i+1} \sqrt{i+1} g$, where we have noted $g=g_{0,1}$. This yields the following coupling Hamiltonian

\begin{equation}
	\label{eqn:coupling_hamiltonian}
	H_I/\hbar = \sum_{i}^{M-2} g_{i,i+1} (\ad\proj{i,i+1} + a\proj{i+1,i}).
\end{equation}
All the experiments reported in this chapter will be carried out in the limit where the detuning $\Delta=\omega_{01} - \omega_r$ between the transmon and resonator transition frequencies is much larger than the coupling constant $g$, the so-called dispersive limit. Approximate expressions for the interaction Hamiltonian can then be derived by applying perturbation theory to various orders. In this work we will be both interested in a ``low-energy'' limit in which the average number of photons in the resonator $\bar{n}$ is small compared to the critical photon number $n_{\rm crit}=(\Delta)^2/(4 g^2)$, and also in the ``high-energy'' limit where this is no longer true. In this case a more refined treatment is needed (see section \ref{sec:back-action_of_a_driven_kerr_resonator_on_a_qubit}). In the low-energy limit $\bar{n} \ll n_{crit}$, the simplest second-order perturbation theory yields an effective Hamiltonian which, restricted in the $\{\ket{0},\ket{1}\}$ subspace, writes:
\begin{equation}
\label{eqn:dispersive_hamiltonian}
	H_{int}= \hbar ( \chi \sigma_z + \bar{s} ) a^\dagger a,
\end{equation}
where $\sigma_z$ is the usual Pauli matrix (such that the qubit Hamiltonian writes $H_q=-\hbar \omega_{01}/2 \sigma_z$), $\chi = \chi_{01} - \chi_{12}/2$, $\chi_{01}=g^2/\Delta$, $\chi_{12}=g_{12}^2/(\omega_{12} - \omega_r)$ and $\bar{s}=- \chi_{12}/2$~\shortcite{Koch2007}.

In summary, including the dissipative terms, the qubit drive, and the nonlinear resonator pump, we get the following master equation
\begin{equation}
	\label{eqn:system_masterequation}
	\dot\rho = \sL\rho = - \frac{i}{\hbar}\comm{H_I}{\rho} + \sL_q\rho + \sL_r\rho 
\end{equation}
that describes the dispersive interaction between a superconducting qubit and a KNR.



\section{Semiclassical dynamics of the nonlinear resonator} 
\label{sec:semiclassical_dynamics_of_the_nonlinear_resonator}

In this section we temporarily leave the qubit aside and study the dynamics of the pumped KNR alone described by \eqref{eqn:NLO_masterequation}. We identify the bistability region by calculating the classical response of the KNR to the pump. We then focus on calculating the switching probability of the resonator between its two dynamical states as a function of the amplitude and frequency of the microwave pump pulse. Numerical simulations are finally compared to measurements.

\subsection{Classical response to a pump} 
\label{sub:classical_response_to_a_pump}
\begin{figure}[t]
\begin{center}
\includegraphics[width=8cm,angle=0]{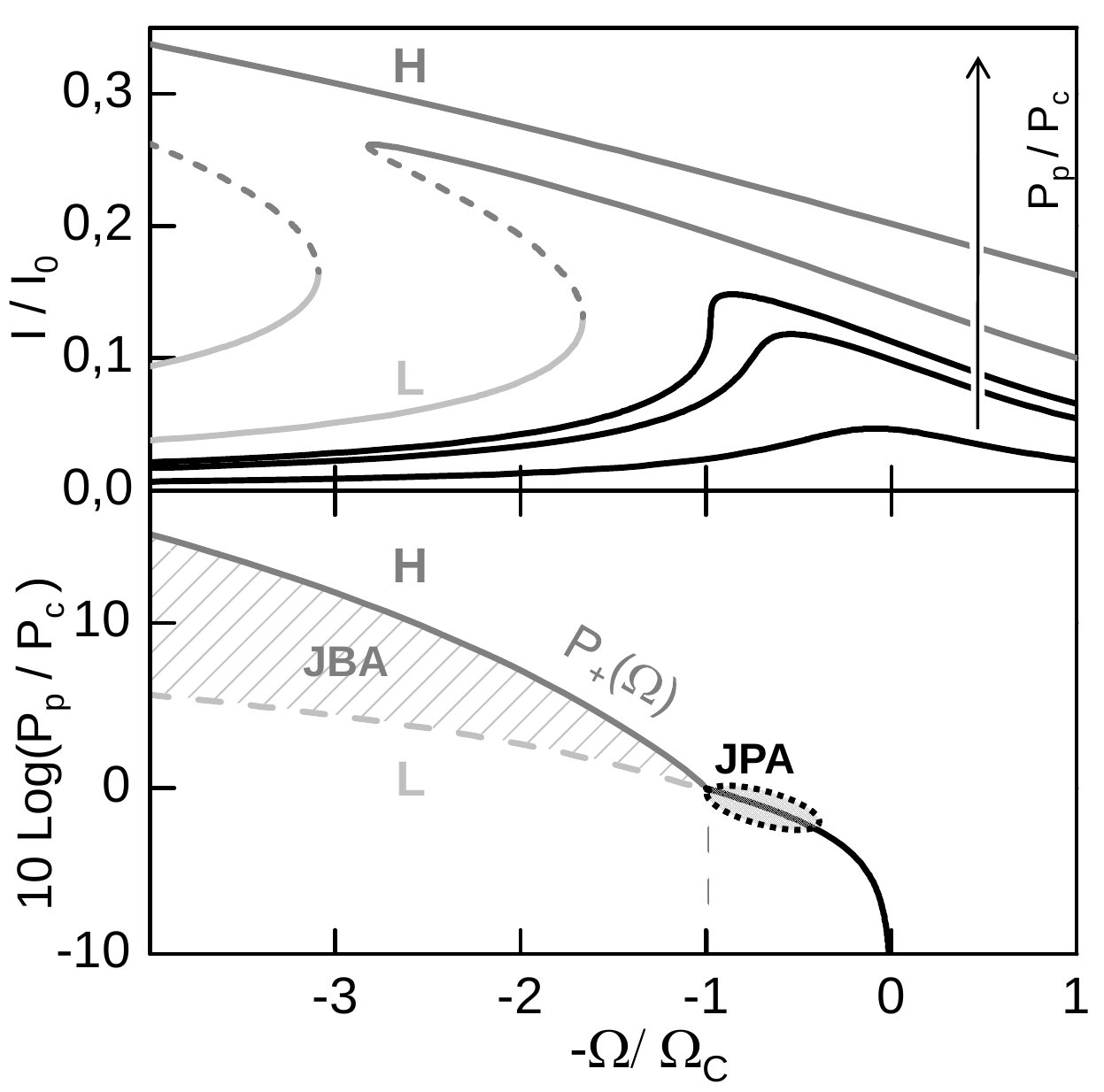}
\end{center}
\caption{Classical response of a KNR to a pump microwave signal. (top) Calculated maximum current $I$ through the junction as a function of the reduced detuning $\Omega$ for various pump powers in critical power ($P_{\rm c}$) units. (bottom) Stability regions of the  oscillator driven at frequency $\Omega$ and power $P_{p}$. Region $L$ ($H$) corresponds to low (high) oscillation amplitude, whereas in the hatched region both solutions are metastable. In the bistable region ($\Omega>\Omega_C$) the KNR acts as a Josephson Bifurcation Amplifier (JBA), whereas for $\Omega<\Omega_C$ the KNR acts as a Josephson Parametric Amplifier (JPA) with maximum gain close to the critical point $(\Omega_C,P_C)$.}
\label{figphasediagram}
\end{figure}

The steady-state complex amplitude $\alpha$ of the intracavity field in the classical limit is obtained from \eqref{eqn:NLO_masterequation}~\shortcite{dykman_theory_1979,drummond_quantum_1980,Yurke2006}
\begin{equation}
i \left(\Omega \frac{\kappa}{2} \alpha + K |\alpha |^2 \alpha \right) + \frac{\kappa}{2} \alpha = -i \epsilon_p,
\label{alpha}
\end{equation}
where $\Omega=2 Q (1 - \omega_{p} / \omega_r)$ is the reduced detuning of the pump and $\epsilon_p$ its reduced amplitude. This equation can be solved numerically, yielding the amplitude of the oscillating current $I=\sqrt{\hbar / \pi Z_0} \omega_r |\alpha| $ passing through the Josephson junction. This is illustrated in Fig.~\ref{figphasediagram} for typical sample parameters. At low drive amplitude $\epsilon_p$, the response is a Lorentzian centered around $\Omega=0$ as for a linear resonator. When the drive amplitude increases, the resonance shifts downwards and shows a sharpened response for some frequencies. In this regime, small variations in the pump amplitude produce a large change in $\alpha$, indicating that the KNR acts as a Josephson parametric amplifier (JPA)~\shortcite{castellanos-beltran_widely_2007}. For some critical drive power $P_c$, the slope $| d\alpha / d \epsilon_p |$ becomes infinite at $\Omega=\Omega_c=\sqrt{3}$. For $P_p>P_c$, two stable solutions $L$ and $H$ with different oscillation amplitude $\alpha$ exist. In this bistable regime, occurring only for $\Omega>\Omega_c=\sqrt{3}$, the transition from $L$ to $H$ occurs abruptly when ramping up the pump power at a value $P_+(\Omega)$ called the bifurcation threshold and is hysteretic~\shortcite{landau_mechanics}. This regime is called the Josephson Bifurcation Amplifier (JBA)~\shortcite{Siddiqi2006a}. Fig.~\ref{figphasediagram} summarizes these properties.

\subsection{Fluctuations: a qualitative discussion} 
\label{sub:fluctuations}

Fluctuations are essential to understand the physics of a pumped KNR. When the resonator is biased in the JPA regime, fluctuations can be either amplified or de-amplified depending on their phase relation with the pump. This results in a phase-dependent noise of the parametric amplifier. When the resonator is biased in the JBA regime, the switching from the L to H state is a stochastic process because of fluctuations which make the dynamics of the intracavity field differ from one experiment to the other even with identical biasing conditions. A quantitative account of fluctuations is then needed in order to determine the probability that the resonator switches from L to H, as well as the switching rate.

In the JPA regime, fluctuations can be treated by linearizing the equations of motion around the steady-state solution $\alpha$ [see \eqref{alpha}]. The input-output relations are then solved showing how input fluctuations are amplified or de-amplified at the output, with a gain that depends on the pumping conditions. This method has been used for instance in~\shortcite{drummond_quantum_1980,PhysRevA.32.2887,Yurke2006} to derive explicit expressions for the phase-dependent gain and noise properties in the JPA regime, and a complete quantum theory was developed in~\shortcite{PhysRevA.83.052115}. In the classical version of this problem, such analysis as well as the effect of switching events on the gain and on the power spectra was considered in~\shortcite{dykman_theory_1979,PhysRevE.49.1198}. Calculation of switching rates in the JBA regime (which is our main interest here) requires going beyond a linearized analysis of the problem. Approximate theoretical expressions for the transition rate~\shortcite{dykman_theory_1979,Dykman_Smelyanskiy_JETP67} can be obtained by the so-called real-time instanton technique applied to the full non-linear system. Using the critical slowdown of the KNR dynamics close to the bifurcation threshold the problem can be mapped onto the Fokker-Planck equation describing the escape of a classical Brownian particle in a 1D cubic potential \shortcite{dykman_fluctuations_1980,PhysRevA.33.4462,PhysRevE.75.011101}. At this point one can obtain expressions for the switching rate using standard techniques. A very accessible account of this approach is given in~\shortcite{Vijayaraghavan2008}. However these approaches only treat the case in which the pump of the KNR is kept at a constant amplitude or varied very slowly compared to the KNR dynamics. Experimentally, the switching probability is measured with microwave pulses that can induce transients of the intra-resonator field when the rate at which the pump amplitude is varied is comparable to the ringing time $\kappa^{-1}$ of the KNR. For the parameters of our experiments (see sections \ref{sec:cba_readout} and \ref{sec:back-action_of_a_driven_kerr_resonator_on_a_qubit}), the duration of the pulses indeed requires to deal with transients. 

For that reason, we rely here on numerical simulations of the KNR switching from L to H when it is subject to a microwave pump pulse. This approach can treat arbitrary dynamics of the pump amplitude and naturally takes transients into account. In the following, we compare measurements of the switching curves of a KNR whose parameters are known with great accuracy to the simulations.


\subsection{Experimental implementation} 
\label{sub:experimental_implementation}

\begin{figure}[t]
\begin{center}
\includegraphics[width=12cm,angle=0]{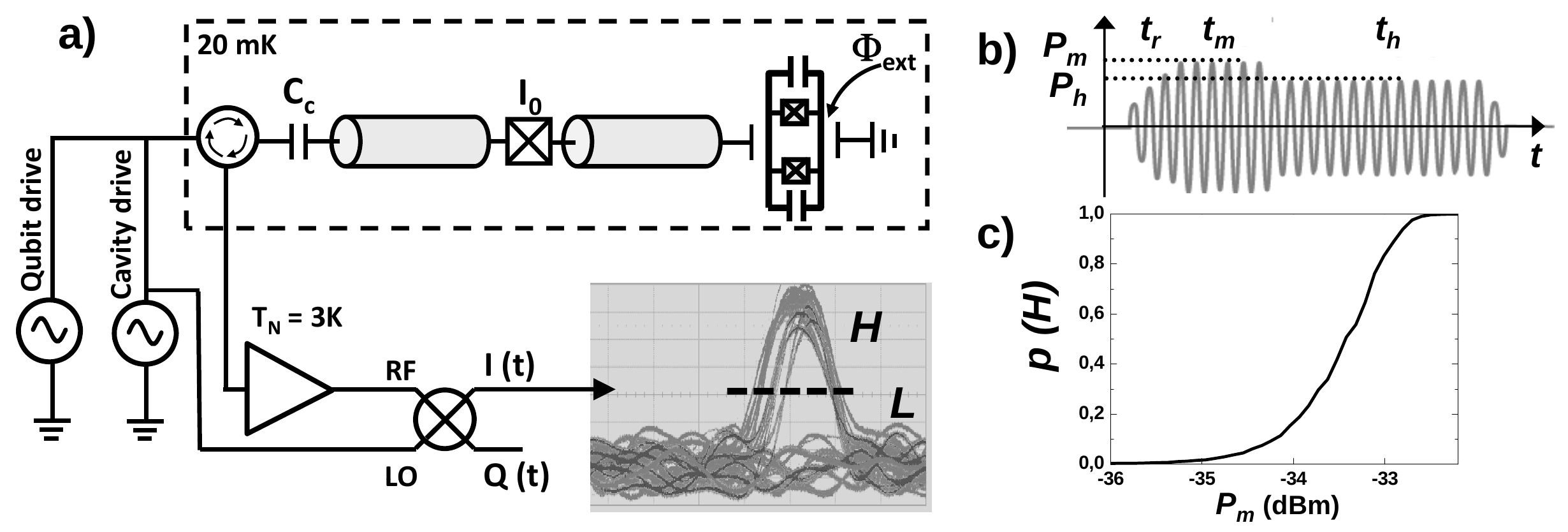}
\end{center}
\caption{(a) Experimental setup for S-curve measurement. A microwave pulse at frequency $\omega_m$ is sent onto the KNR. After reflection onto the KNR, the pulse is amplified and demodulated at room-temperature. The oscillogram shows filtered $I(t)$ traces corresponding to switching and non switching events for $P_m$ close to the bifurcation. The dashed line is a threshold defined to extract the switching probability $p_s$ and acquire S-curves. (b) Temporal profile of the microwave pulse: the microwave amplitude is ramped linearly up to power $P_m$ in a time $t_r$, kept constant for a measurement time $t_m$, and reduced to power $P_h$ during a {\it hold} time $t_h$. (c) Typical S-curve $p_s(P_m)$.}
\label{figscurves}
\end{figure}

The experimental setup is sketched on Fig.~\ref{figscurves}~a). The KNR is capacitively coupled to a transmon qubit. In the experiments reported in this section, the qubit is biased at a flux such that its resonance frequency is much lower than the resonator's to avoid perturbing the KNR dynamics. The KNR has a fundamental frequency $\omega_r / 2\pi =$ 6.4535 GHz (in the absence of qubit), quality factor $Q = 685 $ and Josephson junction critical current $I_{0} = 720 $ nA. These parameters are determined experimentally in separate measurements. Microwave pulses at frequency $\omega_m$ (corresponding to reduced detuning $\Omega$) are sent to the KNR with a shape shown in Fig.~\ref{figscurves}~b). The microwave amplitude is first ramped linearly in $t_r$ from $0$ to the maximum amplitude corresponding to a power $P_m$. The power is then kept constant for a time $t_m$ (effective measurement time) during which the resonator may or may not switch from $L$ to $H$. The power is finally decreased to a hold value $P_h$ at which the resonator state cannot change due to hysteresis, for a {\it hold} time $t_h$ long enough (typically $1\mu s$) to allow good discrimination at room temperature of $L$ and $H$~\shortcite{Siddiqi2006a}. After the pulse is reflected by the KNR, it is amplified, and the two quadratures $I$ and $Q$ are measured by homodyne detection at room temperature. For a power $P_m$ close to the bifurcation threshold, the time traces belong to two clearly resolved families of
trajectories corresponding to oscillator states $L$ and $H$ [see Fig.~\ref{figscurves}~a)]. By defining a measurement threshold [dashed line in Fig.~\ref{figscurves}~a)] and repeating the measurement sequence, we can extract the probability $p_s$ to find the resonator in state $H$ as a function of the drive parameters $P_m$ and $\Omega$. We thus obtain so-called S-curves as displayed in Fig.~\ref{figscurves}~c).

\subsection{Comparison to numerical simulations} 
\label{sub:comparison_simulations}

\begin{figure}[h]
\begin{center}
\hspace{0mm}
\includegraphics[width=12.5cm,angle=0]{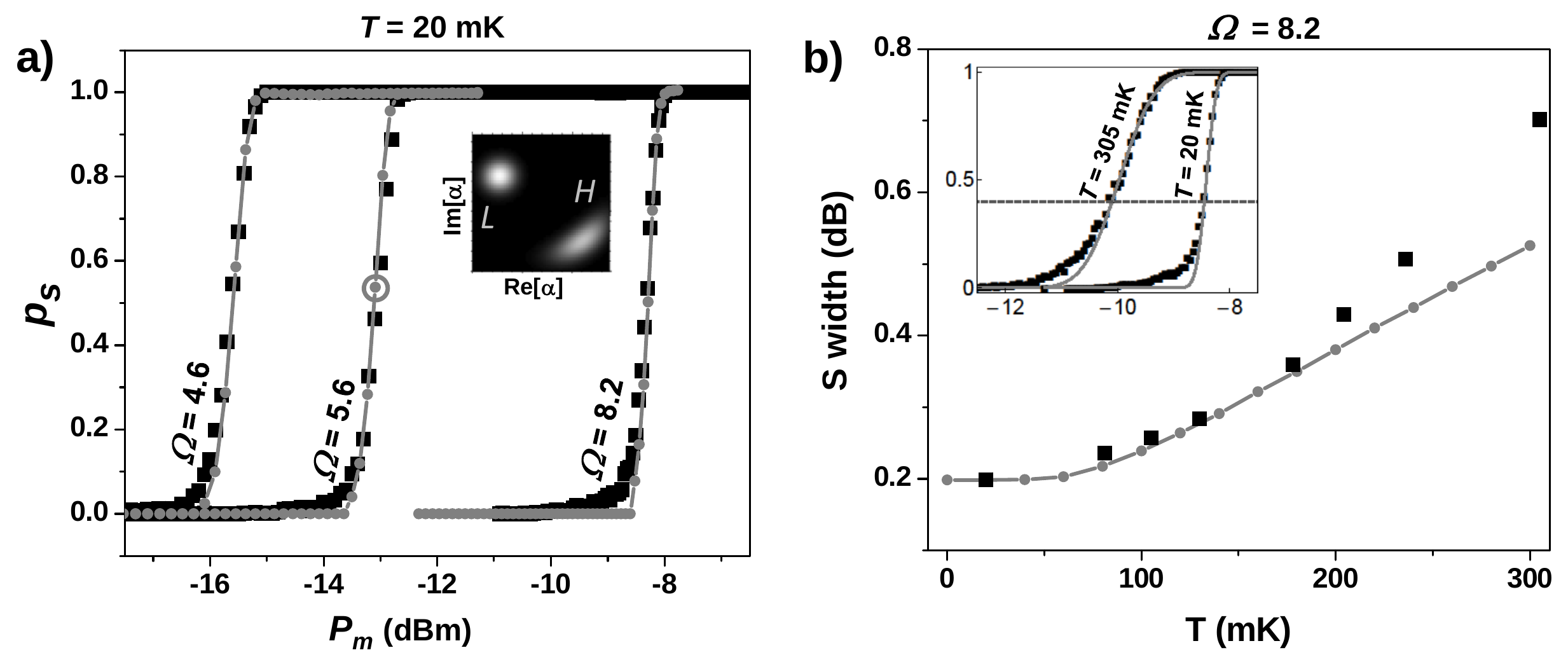}
\caption{Comparison of measured and simulated S-curves.
(a) S-curves at $T$=20 mK for different $\Omega$. Squares are experimental data, dots are simulated data (the grey line is a guide to the eye for the latter). Inset: greyscale representation (black=0, white=0.15) of the function $Q(\alpha)=\bra{\alpha} \rho \ket{\alpha}$ used to infer the circled simulated point in the center of the main panel, $\rho$ being the density matrix of the intra-resonator field calculated by numerical integration of the master equation.
(b) Temperature dependence of the S-curves widths for $\Omega=8.2$. Dots linked by a line are obtained by fitting the simulated S-curves with an $erf$ function (see text). Squares are obtained by fitting the upper part of the experimental S-curves as shown in the inset.
}
\label{fig-hockey}
\end{center}
\end{figure}

Numerical simulations of the S-curves are performed as indicated in the following. The master equation~\pref{eqn:NLO_masterequation} is integrated in the Fock state basis, starting with the resonator initialized in the vacuum state (or in a thermal state at temperature $T$ for finite temperature calculations), and with a time-dependent drive amplitude $\epsilon_p (t)$ taken in exact correspondence to the experimental pulse profile. At the end of the hold time $t_H$, the function $Q(\alpha)=\bra{\alpha} \rho \ket{\alpha}$ ($Q$ distribution~\shortcite{walls_quantum_2008}) is computed from the density matrix $\rho$. When the pulse amplitude approaches the bistability threshold, the $Q$ distribution shows two well-separated regions [see inset of Fig.~\ref{fig-hockey}~a)] corresponding to resonator states $L$ and $H$. The probability $p_S$ is then given by the integrated weight of the high-amplitude part of the $Q$ distribution [see inset of Fig.~\ref{fig-hockey}~a)]. The temperature of the bath can be easily varied as described in~\pref{eqn:NLO_masterequation}. If the bath is at zero temperature, only quantum fluctuations are present. 

In Fig.~\ref{fig-hockey}~a) we first show three S-curves measured for different $\Omega$ with a slow pulse $t_r=300$~ns and $t_m=2\,\mu s$, together with the corresponding simulations rescaled by a global attenuation factor that we have adjusted to fit the data. We first note that the position of the three S-curves is perfectly reproduced by the simulation, indicating that the parameters of the simulation are compatible with our sample. The overall shape of the S-curves is also well reproduced by the simulation, despite a structure at low $p_s$ which is present in the experiment but not in the simulations and that we attribute to the residual effect of the transmon qubit on the KNR\footnote{Indeed, in these measurements the transmon was deliberately biased at low frequency (around or below $4$~GHz to shift it far away from the KNR frequency), resulting in a finite thermal population even at the lowest temperature $T=20$~mK of our cryostat. Excited states of the transmon correspond to a S-curve shifted at a lower $P_m$, and thus the observed S-curve should be seen as a sum of several S-curves weighted by the various excited state populations causing the structured S-curve foot observed in Fig.~\ref{fig-hockey}~a). We are unfortunately unable to quantitatively account for this structure since in this experiment we don't accurately know either the transmon's frequency or its effective temperature.}. Apart from this structure at the S-curve foot, the agreement between the simulations and the experiment is remarkable at all $\Omega$ with the global attenuation as the only adjustable parameter. Since these simulations were realized assuming zero temperature, this constitutes a strong indication that the width of the experimental S-curve is essentially dominated by quantum fluctuations. Note that we expect that the transitions between the two states are not due to tunneling but take place over the effective potential barrier, in the process termed 'quantum activation' in~\shortcite{PhysRevA.73.042108}.

We also measure the temperature dependence of the S-curve width (determined taking into account only the large-$p_s$ part of the curve) for similar pulse parameters and compare it to the simulations [see Fig.~\ref{fig-hockey}~b)]. The agreement is quantitative up to $T \approx 200$~mK. Deviations are observed between $200$ and $300$~mK which might be caused by a decrease of the KNR quality factor due to thermally excited quasiparticles, or by thermal excitation of the qubit. We observe that the crossover from the thermal to the quantum regime in both the experiment and the simulations occurs around $100$~mK. This is significantly lower than the theoretical prediction of $\hbar \omega_r / 2 k_B$ which is around $150$~mK for our parameters~\shortcite{PhysRevA.73.042108,vijay_invited_2009}. More detailed investigations are needed to address this question.


\section{High-fidelity qubit readout by a nonlinear resonator} 
\label{sec:cba_readout}

We now describe how the bistability of the nonlinear resonator combined with its dispersive interaction with the qubit [see \eqref{eqn:dispersive_hamiltonian}] can be exploited to yield a high-fidelity qubit state readout~\shortcite{mallet_single-shot_2009}.

\subsection{Principle} 
\label{sub:principle_of_the_readout_by_cba}
The underlying principle of this readout method can be understood with the dispersive qubit-resonator interaction Hamiltonian \pref{eqn:dispersive_hamiltonian}, valid provided $|\Delta| \ll g$ and $\bar{n} \ll n_{crit}$ as explained in the first section. In this regime, the resonator frequency takes the qubit-state dependent value $\omega_{r,i} = \omega_r + \bar{s} \pm \chi$. Detecting this frequency change sufficiently fast before the qubit relaxes by spontaneous emission is the key challenge of all readout methods in circuit QED. This is usually done by sending a microwave pulse close to the resonance frequency and measuring the phase of the signal transmitted through or reflected by the resonator~\shortcite{Blais2004}, which bears information on the resonator frequency and therefore on the qubit state. One outstanding feature of this dispersive scheme is that it perturbs the qubit state minimally as evidenced by the fact that the interaction Hamiltonian commutes with $\sigma_z$, implying that the readout should be quantum non-demolition (QND)~\shortcite{RevModPhys.68.1}.

However this readout method suffers from a limited signal-to-noise ratio (SNR) when applied to a simple linear resonator.
Indeed the mean photon number $\bar{n}$ in the resonator should be kept much lower than $n_{\rm crit}=\Delta^2 / \lp 4 g^2 \rp $
to avoid spurious qubit excitation~\cite{boissonneault_nonlinear_2008,Boissonneault2009}, which limits the measurement power. Such a small signal must be amplified for detection but state-of-the-art cryogenic amplifiers have a noise temperature $T_{\rm N} \approx$ 3 K so that the added noise is much larger than the signal. Typical figures are $g/\Delta \sim 1/10$, so that $n_{\rm crit} \sim 25$ whereas the amplifier adds a noise corresponding to $25$ photons.
Averaging during a time $t_{\rm m}$ improves the SNR, but the longer $t_{\rm m}$, the higher the probability that the qubit (characterized by a decay time $T_1$ of the order of the microsecond) relaxes yielding an incorrect readout. There thus exists an optimal $t_{\rm m}$ yielding an optimum discrimination between both qubit states which can be shown to be of order $T_1$~\shortcite{gambetta_protocols_2007}.
These constraints have up until now made it impossible to reach the readout efficiencies required for quantum information applications, even though quantum state tomography is still possible by performing ensemble averages~\shortcite{filipp_two-qubit_2009}.
In the following, we show that turning the resonator into a KNR makes it possible to improve the visibility of the measurement without compromising either its QND character or the qubit coherence properties.

\begin{figure}[h]
\begin{center}
\hspace{0mm}
\includegraphics[width=12cm,angle=0]{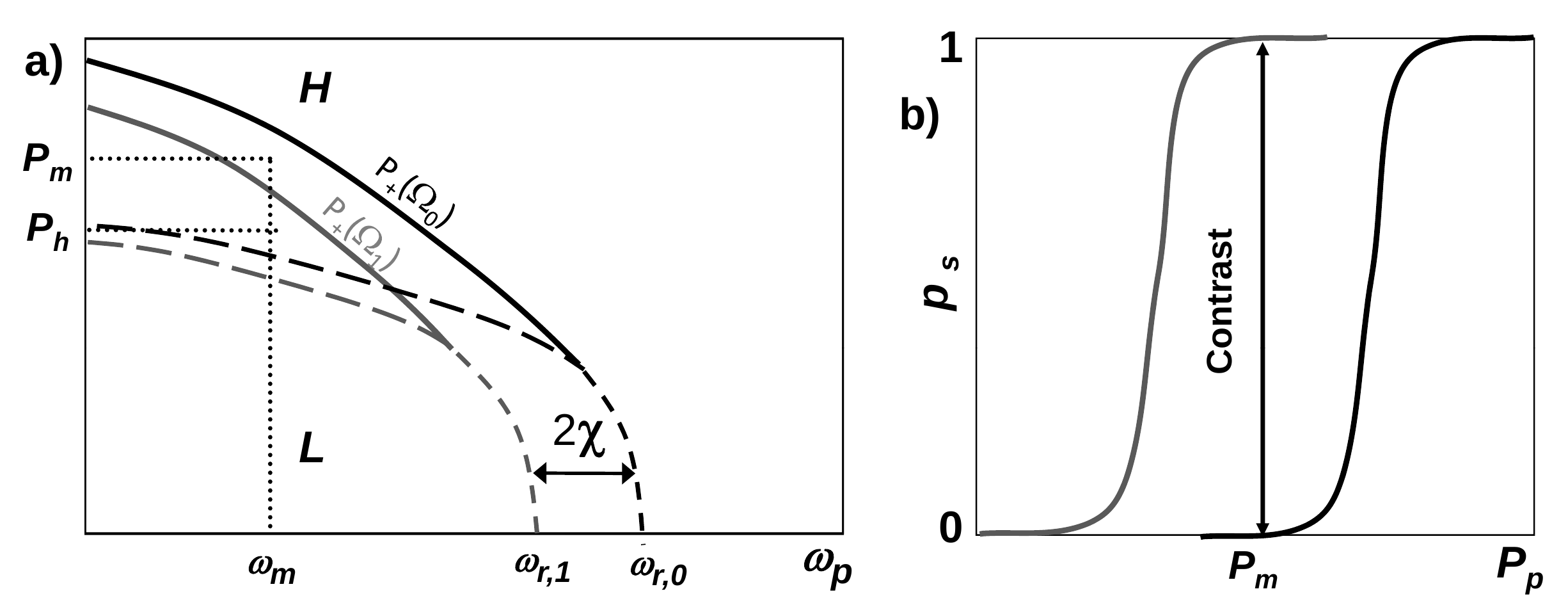}
\caption{Principle of JBA readout.
(a) Sketch of the stability diagrams of the NL resonator dispersively coupled to a transmon prepared in state $\ket{0}$ (black) or $\ket{1}$ (grey).(b) Sketch of the probability $p_s$ to measure the resonator in state $H$ for qubit initially in $\ket{0}$ (black) or $\ket{1}$ (grey).
}
\label{fig-readout-principle}
\end{center}
\end{figure}

Taking advantage of the bistability, a KNR dispersively coupled to a qubit can be turned into a sample-and-hold qubit state detector with a large SNR~\shortcite{Siddiqi2006,boaknin_dispersive_2007}. Due to the dispersive coupling \pref{eqn:dispersive_hamiltonian} the reduced pump detuning $\Omega_{i}$ and the bifurcation threshold $P_+(\Omega_{i})$ now depend on the qubit state. Sending microwave pulses at a frequency and a power $P_m$ such that $P_+(\Omega_{1}) < P_m < P_+(\Omega_{0})$ then maps the
resonator states $L$ or $H$ onto qubit states $\ket{0}$ or $\ket{1}$ as shown in Fig.~\ref{fig-readout-principle}~a)~\shortcite{mallet_single-shot_2009}. The pulses have a temporal shape identical to those described in the previous section [see Fig.~\ref{figscurves}~b)], allowing perfect single-shot discrimination of the two resonator states as already discussed. The key advantage of this readout method over readout with a linear resonator is that the outcome of the measurement is totally insensitive to relaxation of the qubit if the qubit relaxes during the {\it hold} part of the readout pulse. However qubit relaxation occurring before the resonator switches causes measurement errors as we will see below. This approach results in potentially large readout fidelity provided the separation between the S-curves corresponding to both qubit states
is large compared to their width [see Fig.\ref{fig-readout-principle}~b)]. Such an rf-switching scheme has been implemented successfully to readout a quantronium qubit~\shortcite{Boulant2007,siddiqi_dispersive_2006},
a flux qubit~\shortcite{Lupascu2006} and a transmon qubit~\shortcite{mallet_single-shot_2009}, as described in the next section. We finally note that two recent experiments use nonlinearity to achieve high-fidelity qubit readout in a different way: Vijay et al.~\shortcite{PhysRevLett.106.110502} have used a JPA as a first-stage amplifier for linear dispersive readout and in this way have observed quantum jumps of a transmon qubit, while Reed et al.~\shortcite{reed_high_2010} use the nonlinearity of the cavity inherited by coupling to the qubit to obtain qubit-state dependent signal in the ultra-strong driving regime~\shortcite{Boissonneault2010,Bishop2010}.


\subsection{Implementation of the JBA readout of a transmon} 
\label{sub:implementation_of_cba}

\begin{figure}[h]
\begin{center}
\hspace{0mm}
\includegraphics[width=8cm,angle=0]{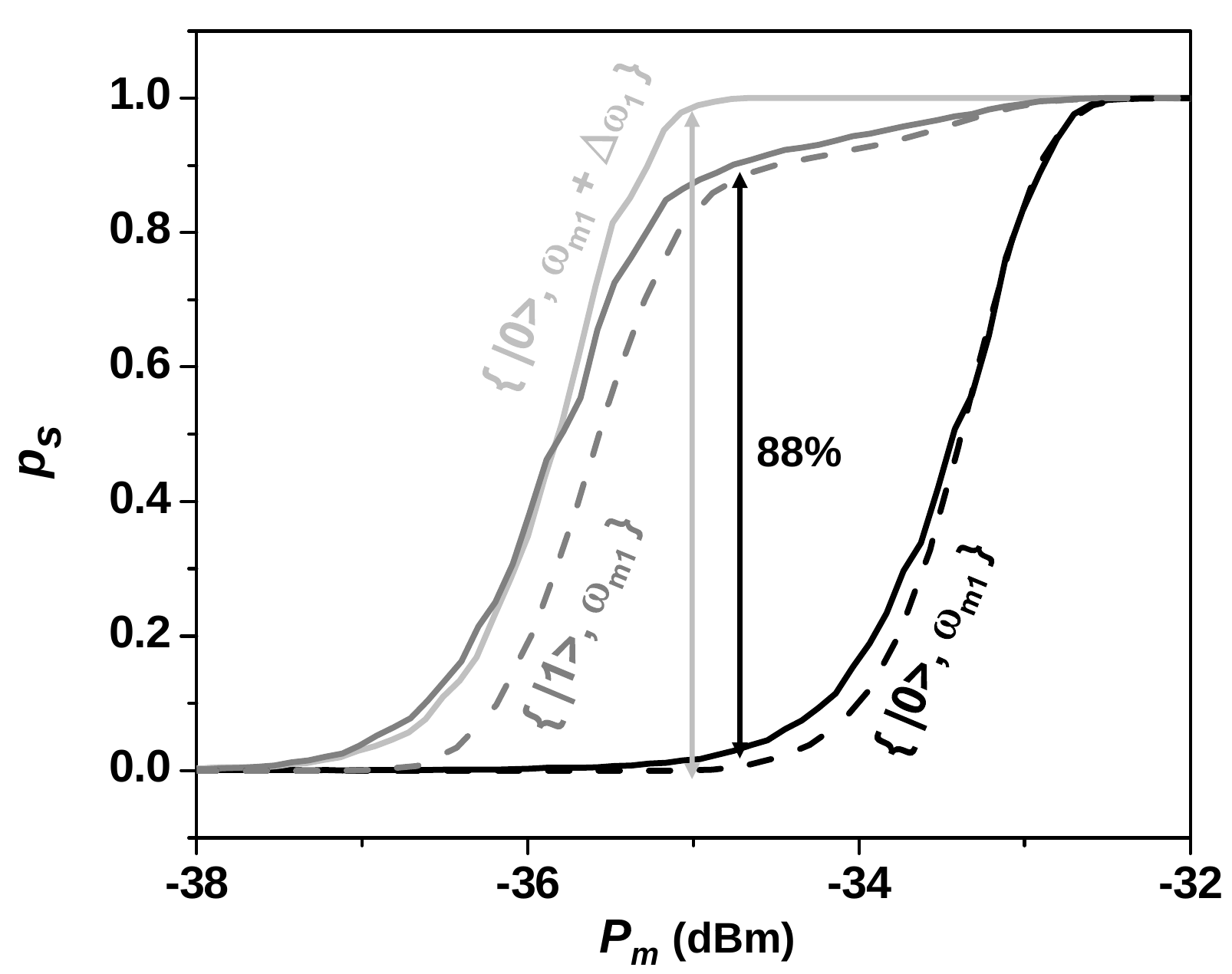}
\caption{Readout contrast: experiment and simulation. S-curves for qubit prepared in $\ket{0}$ (black) or $\ket{1}$ (dark grey) with $\omega_m/2\pi = 6.437 \;\mathrm{GHz} \equiv \omega_{m1}/2\pi$. Full lines are data and dashed lines are simulations. The black arrow shows the maximum experimental contrast. The lighter grey full line is the S-curve obtained with qubit prepared in $\ket{0}$ with the same parameters as for the solid lines except that $\omega_m= \omega_{m1} - \Delta \omega_1$ (see text). The grey arrow shows the expected contrast in absence of relaxation.
}
\label{fig-readout-implementation}
\end{center}
\end{figure}

The experimental setup is the same as shown in Figs.~\ref{figqubitKNR} and~\ref{figscurves}. In contrast to the measurements reported in section~\ref{sec:semiclassical_dynamics_of_the_nonlinear_resonator}, the transmon qubit is now biased at a flux such that its frequency is closer to the KNR so that the dispersive coupling constant $\chi$ becomes sizeable, a necessary requirement for readout with high contrast. In all the measurements reported here the qubit is operated at negative
detunings $\Delta=\omega_{01} - \omega_r$, larger in absolute value than $g/2\pi=44$~MHz. Readout pulses [Fig.~\ref{figscurves}~b)] of frequency $\omega_m$ and power
$P_m$ are sent to the KNR, allowing to measure S-curves as explained in~\ref{sub:experimental_implementation}.

Figure~\ref{fig-readout-implementation} shows S-curves obtained at $- \Delta / 2\pi=380$ MHz with $\omega_m/2\pi = 6.437 \;\mathrm{GHz} \equiv \omega_{m1}/2\pi$ for both qubit states. The $\ket{1}$ state was prepared with a $\pi$ pulse applied just before the readout. The maximum separation, or best contrast, at that working point is 88 $\%$. Also shown is a S-curve with the qubit prepared in $\ket{0}$ and measured in the same conditions but with $\omega_m= \omega_{m1} - \Delta \omega_1$. The detuning $\Delta \omega_1/2\pi=4.1$ MHz is chosen such that the corresponding S-curve matches best the S-curve with qubit prepared in $\ket{0}$ and $\omega_m=\omega_{m1}$. Thus, $\Delta \omega_1$ yields a direct measurement of the cavity pull and is in good agreement with the theoretical value of $2\chi/2\pi=4.35$ MHz calculated for these parameters (see~\ref{sub:coupling}).

However comparing the grey and the light grey S-curves in Fig.~\ref{fig-readout-implementation} we note a discrepancy when $p_s$ approaches unity. This is due to relaxation of the qubit state $\ket{1}$ during the time the resonator energizes and choses its state. Quantitative analysis (not presented here) shows that this time is of the order of $T_1/10 \approx 40$ ns with our circuit parameters. Although the consequences are much less dramatic than in linear circuit-QED, the JBA readout fidelity remains limited by finite $T_1$. To emphasize this point we estimate that the contrast without relaxation would be 97$\%$ (gray arrow on Fig.~\ref{fig-readout-implementation}).

We also present in Fig.~\ref{fig-readout-implementation} the results of a simulation of the complete master equation~\pref{eqn:system_masterequation}, taking into account the measured value of the qubit relaxation time $T_1$ and of the qubit-resonator coupling constant. Even though the shift of the S-curves between states $\ket{0}$ and $\ket{1}$ is slightly less in the simulation than in the experiment, the overall agreement with the experimental data is remarkable with the global attenuation of the measurement line as the only adjustable parameter. In particular the maximum contrast predicted by the simulation is precisely the one found in the experiment. This demonstrates that despite the complexity of our circuit (a multilevel system coupled to a KNR) it is possible to obtain a quantitative and detailed description.

In the next subsection we show how we can partially get rid of the relaxation to improve the fidelity of the readout, and we finally present a general figure of merit of the JBA transmon readout.

\subsection{Performance and limitations of the transmon readout by a JBA} 
\label{sub:performances_of_cba}

\begin{figure}[h]
\begin{center}
\hspace{0mm}
\includegraphics[width=12cm,angle=0]{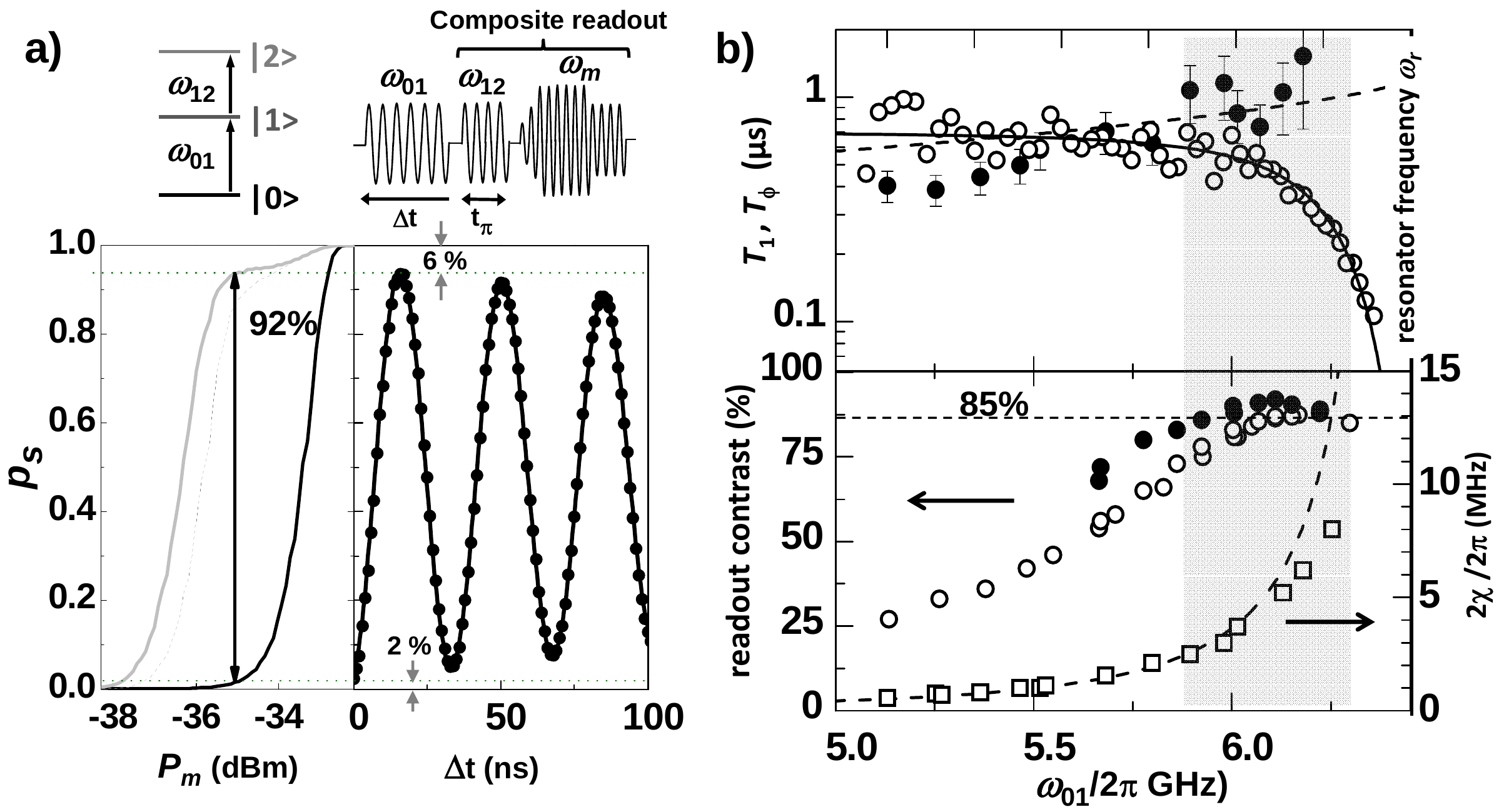}
\caption{Performances of JBA readout.
a) Best single shot readout contrast. Left: S-curves with qubit prepared in $\ket{0}$ (black), $\ket{1}$ (dotted grey) and $\ket{2}$ (solid grey). Right: Rabi oscillations measured with the composite readout (dots). The solid line is a fit by an exponentially damped sine curve (peak-to-peak amplitude 94$\%$ and decay time 0.5 $\mu$s).
b) Trade-off between qubit coherence and readout fidelity. Top: experimental relaxation time $T_1$ (dots) and dephasing time $T_\phi$ (open circles) of the qubit as a function of $\omega_{01}$. The solid line is the value of $T_1$ obtained by adding to the expected spontaneous emission through the resonator
a relaxation channel of unknown origin with $T_1=0.7~\mu$s. The dashed line is the pure dephasing time $T_\phi$ corresponding to a $1/\omega$ flux noise with an amplitude set to 20 $\mu \Phi_0/\sqrt{\rm Hz}$ at 1 Hz.
Bottom, left axis: Readout contrast with (dots) and without (open circles) transfer from state $\ket{1}$ to $\ket{2}$. Bottom, right axis: measured (open
squares) and predicted (dashed line) cavity pull $2\chi$. The grey area highlights the detuning region where the readout contrast is higher than 85 $\%$.
}
\label{fig-readout-performances}
\end{center}
\end{figure}

To reduce the effect of relaxation and improve the readout contrast we use a technique analogous to electron shelving in atomic physics and already used with other Josephson qubits~\shortcite{martinis_rabi_2002} : we transfer state $\ket{1}$ into the next excited
state $\ket{2}$ with a resonant $\pi$-pulse just before the readout pulse. This has two effects on the readout performance. First, it can be shown that the resonator frequency is slightly more shifted by state $\ket{2}$ of the transmon than by state $\ket{1}$, yielding a larger S-curve separation between the two qubit states and thus an enhanced readout fidelity. Second, relaxation from state $\ket{2}$ occurs only towards state $\ket{1}$ (which hardly changes the outcome of the readout if $P_m$ is properly chosen), and not directly towards $\ket{0}$ in the transmon, as explained in section \ref{sub:qubit}. Still working at $- \Delta / 2\pi=$ 380 MHz, we obtain the solid grey S-curve on Fig.~\ref{fig-readout-performances}~a) yielding a $92\%$ contrast.
The right panel shows Rabi oscillations between $\ket{0}$ and
$\ket{1}$ obtained with such a composite readout pulse. This contrast,
larger than $90\%$, is in agreement with the width of the S-curves estimated from numerical simulations, with their theoretical displacement,
and with the measured qubit relaxation time. Of the remaining $8\%$
readout error, we estimate that about $6\%$ are due to relaxation
before switching and $2\%$ to residual out-of-equilibrium population
of $\ket{1}$ and to control pulse imperfections.

We now discuss the dependence of the readout contrast and qubit coherence
on the detuning $\Delta = \omega_{01} - \omega_{r}$. Besides acting as a qubit state detector,
the resonator also serves as a filter protecting the qubit against
spontaneous emission into the 50 $\rm{\Omega}$ impedance of the
external circuit~\shortcite{esteve_effect_1986,houck_generating_2007}. The smaller $|\Delta|$,
the stronger the dispersive coupling $\chi$ between the qubit and the resonator, implying
a larger separation between S-curves
curves but also a faster relaxation due to the Purcell effect. We thus expect the contrast to
be limited by relaxation at small $|\Delta|$, by the poor separation
between the S-curves at large $|\Delta|$, and to exhibit a maximum
in between. Figure~\ref{fig-readout-performances}~b) presents a summary of our measurements
of contrast and coherence times. At small $|\Delta|$, $T_{1}$ is in
quantitative agreement with calculations of the spontaneous emission
through the resonator. However it presents a saturation, similarly
as observed in several circuit-QED experiments~\shortcite{Houck2008}. The cavity
pull determined from the S-curve shifts
[cf. Fig.~\ref{fig-readout-implementation}~b)] is in quantitative agreement with the value
of $2\chi$ calculated from the sample parameters. The contrast varies
with $\Delta$ as anticipated and shows a maximum of $92\%$ at $-\Delta/2\pi=$ 380
MHz, where $T_{1}=0.5\,\mu$s. Larger $T_{1}$ can be obtained
at the expense of a lower contrast, and reciprocally. Another important
figure of merit is the pure dephasing time $T_{\phi}$
which controls the lifetime of a superposition of qubit states~\shortcite{Ithier2005}. $T_{\phi}$
is extracted from Ramsey experiments and shows
a smooth dependence on the qubit frequency, in qualitative agreement
with the dephasing time deduced from a $1/\omega$ flux noise of spectral
density set to $20 \mu{\rm{\Phi}}_0/\sqrt{\rm Hz}$
at $1$ Hz, a value similar to those reported elsewhere~\shortcite{wellstood_low-frequency_1987}.
To summarize our circuit performances, we obtained a $400$ MHz frequency
range [grey area on Fig.~\ref{fig-readout-performances}~b)] where the readout contrast is
larger than $85\%$, $T_1$ larger than
300 ns, and $T_{\phi}$ larger than 700 ns.
Further optimization of the JBA parameters
$I_{\rm 0}$ and $Q$ might increase this high-fidelity readout frequency window.

We finally discuss the QND character (or projectiveness) of the JBA readout~\shortcite{RevModPhys.68.1}. Since the dispersive interaction commutes with the qubit Hamiltonian, the qubit initially prepared in an arbitrary superposition should be projected onto the state that is measured.
We define the QND fidelity as $F_{\rm QND} =(P(0|0)+P(1|1))/2$, where $P(i|i)$ ($i=0,1$) is the probability that the qubit state $\ket{i}$ is unchanged after a single measurement~\shortcite{lupascu_quantum_2007}. Josephson Bifurcation Amplifier implemented to readout a flux qubit demonstrated a QND fidelity of 88$\%$~\shortcite{lupascu_quantum_2007}. In the case of a transmon, the data presented in~\shortcite{mallet_single-shot_2009} for two successive readouts yield a QND fidelity of $\approx 40\%$. However this poor QND character was shown to be only due to relaxation occuring during the first readout pulse, and not to a lack of projectivity of the measurement. This is due to the rather long duration of the measurement sequence [Fig.~\ref{fig-readout-principle}~b)]: $t_{\rm s}$ ($t_{\rm h}$) is typically a few tens (hundreds) of nanoseconds so that the overall readout time is a significant fraction of $T_1$. As a comparison, the readout pulse duration in~\shortcite{lupascu_quantum_2007} was a few tens of nanoseconds, which was possible because of the low quality factor ($Q \approx 20$) of the resonator. Using  such a low $Q$ cavity for the JBA would certainly help performing the readout faster but at the price of degrading the fidelity of the readout (broadening of the S-curves) and the coherence times (inefficient Purcell protection).



\section{Backaction of a driven Kerr resonator on a qubit} 
\label{sec:back-action_of_a_driven_kerr_resonator_on_a_qubit}

In the readout method described in the previous section, the KNR is biased very close to its bifurcation threshold. The small frequency shift due to the qubit is then sufficient to push the resonator above the threshold, resulting in a very large difference at the output for the two qubit states. In other words, the nonlinear resonator energized by the pump signal behaves as an active measurement device with some power gain: an on-chip amplifier. This is in strong contrast with the more traditional measurement process in linear circuit QED where the resonator has a purely passive behavior. It then seems very interesting to investigate the quantum backaction of this specific measurement process on the qubit and to compare it to the well-known linear circuit QED case.

Indeed, any measurement in quantum mechanics necessarily induces decoherence in the variable conjugate to the one being measured. Quantum physics imposes the inequality
\begin{equation}\label{eq:ReadoutInequality}
\Gamma_{\phi m} \geq \Gamma_{meas}/2
\end{equation}
between the system's measurement-induced dephasing rate and the measurement rate $\Gamma_{meas}$, stating that the most efficient detector can only measure as fast as it dephases~\shortcite{Clerk2003}. The dispersive readout in cQED with a passive linear resonator has been shown theoretically to reach this quantum limit~\shortcite{Gambetta2008}. In the following, we will investigate whether the qubit readout by an active nonlinear resonator reaches that limit. We will first give a theoretical treatment of this question and then present an experiment performed on the same sample as used in the previous section. In this experiment, we have measured the qubit dephasing rate as a function of the drive amplitude and frequency of the nonlinear resonator, allowing us to test these predictions.

\subsection{Backaction of a nonlinear resonator on a qubit: theory} 
\label{sub:back-action_theory}

\subsubsection{Backaction in linear cQED: theory} 
\label{sub:back-action_of_a_linear_resonator_on_a_qubit}

\begin{figure}[t]
\begin{center}
\hspace{0mm}
\includegraphics[width=4cm,angle=0]{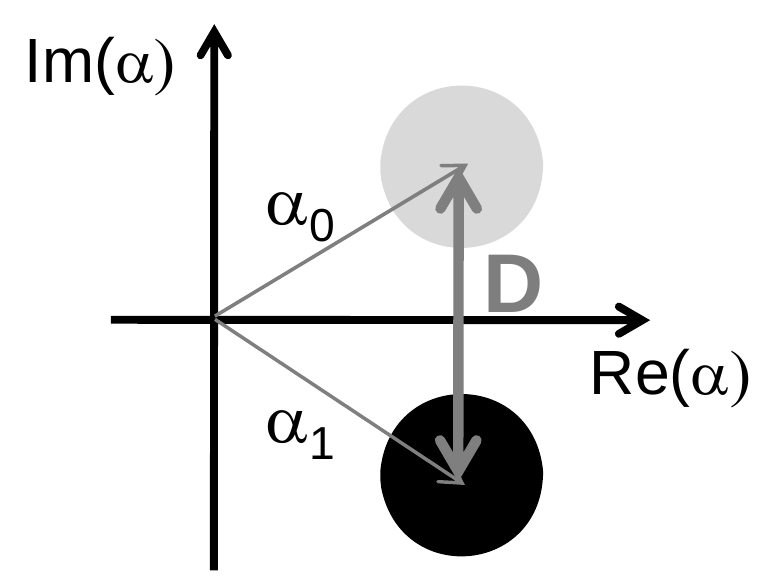}
\caption{Graphical representation of the pointer states $\ket{\alpha_i}$ in phase space and of the distinguishability $D=|\alpha_0 - \alpha_1|$.
}
\label{fig-acstark-pointerstates}
\end{center}
\end{figure}

Before taking into account the Kerr nonlinearity of the resonator, we first review the linear case and consider only two transmon levels for simplicity. Let us consider the qubit prepared in the superposition $c_0\ket{0}+c_1\ket{1}$ ; one then turns on a measurement drive at a frequency $\omega_p$ chosen to be close to the resonator frequency $\omega_r$. The dispersive interaction \eqref{eqn:dispersive_hamiltonian} then leads to an entangled qubit-field state of the form
\begin{equation}
\rho(t) = \sum_{i,j=0,1} c_{i,j}(t) |i,\alpha_i(t)\rangle\langle j,\alpha_j(t)|,
\end{equation}
with $c_{i,i}(t) = |c_i|^2$ as required for a QND interaction. In steady-state, the qubit-state dependent coherent state amplitudes are given by
\begin{equation}
\alpha_{0,1} = \frac{- \epsilon_p}{(\omega_r-\omega_p\pm \chi)-i\kappa/2}
\end{equation}
and are illustrated in Fig.~\ref{fig-acstark-pointerstates}. For large enough distances $D=|\alpha_0 - \alpha_1|$, these two field states can be distinguished in a homodyne measurement and thus act as pointer states for the qubit. The parameter $D$ is called {\it distinguishability} and plays a key role in the following. 

From quantum trajectory theory, the optimal rate at which this  measurement extracts information about the qubit state is found to be $\Gamma_{meas} = \kappa D^2$ and is proportional to the intracavity photon number $\bar{n}$~\shortcite{Gambetta2008}. Moreover, the fundamental backaction of the measurement on the qubit is found in the time-dependence of the off-diagonal elements of $\rho(t)$, which, neglecting transients, decay at a rate
\begin{equation}\label{eq:linearbackaction}
   \Gamma_{\phi m} = 2\chi \mathrm{Im}[\alpha_0\alpha_1^*] = \frac{\kappa}{2} D^2
\end{equation}
saturating the inequality \eqref{eq:ReadoutInequality}. This expression was experimentally confirmed by spectroscopic measurements of the qubit linewidth as a function of $\bar n$~\shortcite{Schuster2005,palacios-laloy_experimental_2010}.

The off-diagonal elements of $\rho(t)$ also experience a frequency shift $\delta\omega_a = 2\chi \mathrm{Re}[\alpha_0(t)\alpha_1^*(t)]$, corresponding to the ac-Stark shift. In the limit $2\chi<\kappa$, this shift linearly follows  the intra-cavity photon number $\delta\omega_a \approx 2\chi \bar n$~\shortcite{Schuster2005}. The increase of the qubit linewidth $ \Gamma_{\phi m}$ with intracavity photon number can then simply be understood as photon shot-noise in the ac-Stark shift. Indeed, each number state $\ket n$ entering the pointer states $\ket{\alpha_i}$ is shifting the qubit frequency by a quantity $2\chi n$  resulting in a broadened spectroscopic line. When $2\chi>\kappa$, the photon number states can be resolved spectroscopically leading to number splitting~\shortcite{Dykman1987,Gambetta2006,Schuster2007,PhysRevA.75.042302}.

We note that the classical fields $\alpha_{0(1)}$ in this linear case can be obtained by linearizing the response around an average field $\bar\alpha$ calculated without the qubit signal provided $2\chi\ll\kappa$. Such a linear response theory can still hold for a nonlinear resonator at low pump power and this regime was studied before~\shortcite{Laflamme2011}. Here, we are interested in regimes in which the bias (pump) drive can approach the bifurcation threshold. Close to these points, the susceptibility $| d \alpha / d\Omega|$ --- and hence the gain $| d \alpha / d\epsilon_p|$ --- of the KNR diverges. The range of signal amplitude (the qubit cavity-pull) for which linear response theory holds therefore vanishes.


\subsubsection{Backaction in nonlinear cQED: theory} 
\label{sub:back-action_of_the_kerr_resonator_on_a_transmon_qubit}

We now turn to the backaction of the nonlinear resonator on the transmon qubit, taking into account its multilevel structure. To go beyond linear response, we use the polaron transformation approach developed in \shortcite{Gambetta2008}. This approach was first used for a two-level system dispersively coupled to a linear resonator. To be useful with a nonlinear resonator and a many-level qubit, it must however be adapted in several ways.

First, in the dispersive readout presented in section~\ref{sub:back-action_of_a_linear_resonator_on_a_qubit}, the photons inside the resonator are assumed to have a frequency $\omega_r$.  This holds for a linear resonator because one always measures with a drive that is (quasi-)resonant with $\omega_r$. On the contrary, we saw in section~\ref{sec:cba_readout} that the JBA readout drive is usually detuned from $\omega_r$ by a few tens of megahertz, or equivalently a few percent of the qubit-resonator detuning. This detuning yields a variation on the ac-Stark shift per photon that needs to be accounted for to obtain quantitative agreement between the theory and the experimental data. Meanwhile, the qubit is still Lamb-shifted by vacuum fluctuations at the resonator frequency. This frequency however now depends on the amplitude of the field inside the resonator. Moreover, we are interested in the qubit spectroscopy while the resonator being pumped. We must therefore consider two drives on the system: one far-detuned from (pump) and one quasi-resonant to (spectroscopy) the qubit frequency. Finally, and as stated above, we must go beyond linear response theory. In the following subsections, we develop a theory~\shortcite{boissonneault2011} that accounts for the aforementioned effects.

\subsubsection{Classical field} 
\label{ssub:classical_field}
The first and most important effect to consider is the classical field created by the pump drive. A displacement transformation in the bare frame~\shortcite{Blais2007} and a polaron transformation in the dispersive frame~\shortcite{Gambetta2008} are two approaches that have previously been used to treat this field. The former captures the pump-resonator detuning in the ac-Stark shift, but, since there is only one field $\alpha$, it is within linear response theory. On the contrary, the latter can go beyond linear response but, since the qubit-resonator dispersive approximation is done first, the ac-Stark shift is a function of the qubit-resonator detuning rather than of the qubit-pump detuning.

Here, we first do a polaron transformation on the full Jaynes-Cummings Hamiltonian. The polaron transformation is defined as 
\begin{equation}
	\tP = \sum_{i=0}^{M-1} \proj{i,i} \tD[\alpha_i(t)],
\end{equation}
where $\tD[\alpha_i(t)]$ is a time-dependent field displacement operator~\shortcite{Gambetta2008}. Defining $H' \equiv \tP^\dag H \tP$, we obtain the Hamiltonian in the polaron basis
\begin{equation}
	H' = H_{NL}' + H_d' + H_q' + H_I' + H_\tP \equiv H_0' + H_1' + H_2',
\end{equation}
where $H_\tP \equiv i\dot \tP^\dag\tP = (-i\dproj{\alpha}\ad + i\dproj{\alpha}^*a) - \ImaginaryPart[\proj\alpha\dproj{\alpha}^*]$ accounts for the time-dependence of the polaron transformation with:
\begin{subequations}
	\label{eqn:H_0_H_1_H_2_prime}
	\begin{align}
		H_0' &= \proj{\omega} + \sum_{i=0}^{M-2} g_i\lsb \proj\alpha^*\proj{i,i+1}+\proj{i+1,i}\proj\alpha\rsb + s(\proj{\alpha}), \\
		H_1' &= \sum_{i=0}^{M-2} g_i(\proj{i,i+1}\ad + \mathrm{h.c.}) + \lsb f(\proj{\alpha})\ad + \mathrm{h.c.}\rsb, \\
		H_2' &= \omega_r'(\proj{\alpha})\ada,
	\end{align}
\end{subequations}
where $\omega_r'(\proj{\alpha}) = \omega_r + 2K|\proj{\alpha}|^2 + 3K'|\proj{\alpha}|^4$, $s(\proj{\alpha})$ and $f(\proj{\alpha})$ are functions that are diagonal in the qubit basis and act as identity in the qubit subspace if $\alpha_{i}=\alpha_j$, and $\proj{\alpha}\equiv\sum_{i=0}^{M-1}\proj{i,i}\alpha_i(t)$. The subscript on these three transformed Hamiltonians refer to the number of resonator ladder operators involved. In obtaining these transformed Hamiltonians, we have assumed that the ``classical part'' of the fields $\alpha_i$ is larger than the ``quantum part'', i.e. that $|\alpha_i|^2 \gg \mean{\ada}'_{\rm ss}$, where this last mean value is taken in the steady-state and in the polaron frame. Doing so, we have dropped all terms with more than one resonator ladder operator (${\ad}^2,a^2,\ada a,...$), except diagonal ones ($\ada$). These terms would lead to squeezing-related effects~\shortcite{Laflamme2011}, which we consider to be small here. As shown in \shortciteNP{Gambetta2008}, the polaron transformation applied on $\proj{i,i+1}$ yields 
\begin{equation}
	\proj{i,i+1}' = \proj{i,i+1} \tD^\dag[\alpha_i]\tD[\alpha_{i+1}] = \proj{i,i+1} \tD[\alpha_{i+1}-\alpha_{i}] e^{-i\ImaginaryPart[\alpha_{i+1}^*\alpha_i]}
\end{equation}
Here, we truncate the series at the zeroth order and rather take $\proj{i,i+1}'=\proj{i,i+1}$. While we allow for the pointer states to be different, this simplification imposes the limit
\begin{equation}
	|\alpha_{i+1}-\alpha_{i}|^2 \ll 1/\mean{\ada}'_{\rm ss}
\end{equation}
on their separation. We note that this approximation can hold for a large separation of the pointer states if $\mean{\ada}'_{\rm ss}$ is small in the polaron frame. It is exact in the dispersive regime for a linear resonator and also is a good approximation in the limit of small squeezing. 

Following \shortciteNP{Gambetta2008}, applying the polaron transformation on the dissipative part of the master equation yields
\begin{equation}
	\sD[a']\rho' = \sD[a]\rho' + \sD[\proj{\alpha}] - \frac{i}{2}\comm{i(\proj\alpha^* a-\proj\alpha\ad)}{\rho'} + a\comm{\rho'}{\proj\alpha^*} + \comm{\proj\alpha}{\rho'}\ad.
\end{equation}
The second term of this equation contains measurement induced dephasing, while the third acts as a drive on the resonator and will be cancelled by a proper choice of the $\alpha_i$'s. The two last terms act as lowering operators in the polaron frame and can be neglected if $\mean{\ada}'_{\rm SS}$ is small. In the following steps, we want to diagonalize $H_0'$ --- which will give us the ac-Stark and Lamb shifts --- and choose the displacement operator P so as to eliminate eliminate $H_1'$, which will ensure that the photon population is small in the polaron frame.

\subsubsection{Stark shift} 
\label{ssub:stark_shift}
The effective drive on the qubit $g_i\proj{\alpha}\proj{i,i+1}+\hc$ yields an ac-Stark shift of the qubit frequencies. To compute this shift, we first assume that the field in the cavity can be written as $\proj{\alpha} = \proj{\alpha_p} e^{-i\omega_p t} + \proj{\alpha_s} e^{-i\omega_s t}$. Since $\omega_p$ is far-detuned from the qubit frequency, this drive can be treated dispersively. However, since $\omega_s$ can be resonant with the qubit, this drive must be treated exactly even if $\alpha_s$ is small in practice. Here, we obtain the effect of the pump drive on the qubit by applying a dispersive transformation of the form
\begin{equation}
	\tD = \exp\lsb \sum_{i=0}^{M-2} \xi_i(t) \proj{i,i+1} - \xi^*(t) \proj{i+1,i}\rsb,
\end{equation}
on the Hamiltonian and on the dissipative Linblad terms of the master equation. Considering only the pump drive (since $\proj{\alpha_s}$ is assumed to be small), choosing
\begin{equation}
	\xi_i = \frac{-g\alpha_{p,i}}{\omega_{i+1}-\omega_i - \omega_p},
\end{equation}
and expanding $H_0'' \equiv \tD^\dag H_0'\tD$ to fourth order using Hausdorff's relations, we can show that the off-diagonal terms related to the pump drive in $H_0''$ vanish. We obtain
\begin{equation}
	H'' = H_0'' + H_1'' + H_2'' + H_\tD,
\end{equation}
where
\begin{subequations}
	\begin{align}
		H_0'' + H_{\tD} &= \proj{\omega''(\alpha)} + \sum_{i=0}^{M-2} g_i (\alpha_{s,i} e^{-i\omega_s t} \proj{i+1,i} + \hc), \\
		H_1'' &= \sum_{i=0}^{M-2} g_i(\proj{i,i+1}\ad + \mathrm{h.c.}) + \lsb \lp \proj{f_p(\alpha)} e^{-i\omega_p t} + \proj{f_s(\alpha)} e^{-i\omega_s t}\rp \ad + \hc \rsb \\
		\omega_i''(\alpha) &= \omega_i + S_i^p|\alpha_p|^2 + \frac{1}{4} K_i^p |\alpha_p|^2, \\
		S_i^p &= -(\chi_i^p - \chi_{i-1}^p), \\
		K_i^p &= -4 S_i^p(|\lambda_i^p|^2 + |\lambda_{i-1}^p|^2) \\
		&\quad - (3\chi_{i+1}^p|\lambda_i^p|^2 - \chi_{i}^p|\lambda_{i+1}^p|^2) + 3(\chi_{i-2}^p|\lambda_{i-1}^p|^2 - \chi_{i-1}^p|\lambda_{i-2}^p|^2),
	\end{align}
\end{subequations}
and where $\lambda_{i}^p = -g_i/(\omega_{i+1}-\omega_i - \omega_p)$, $\chi_i^p = -g_i\lambda_i^p$, and $H_2' = H_2''$. We note that contrary to~\shortcite{Gambetta2008} and in agreement with~\shortcite{Blais2007}, the drive frequency $\omega_p$ appears in the dispersive shifts rather than the resonator frequency. This is the correct dependence, and results because we made the polaron transformation before the dispersive approximation.

When obtaining $H_0''$ and $H_2''$, we have assumed that $\alpha_{p,i} \approx \alpha_{p,j} = \alpha_p$. Still, contrary to what is obtained by a single displacement transformation rather than the polaron transformation, the field depends explicitly on the qubit state. Indeed, $\alpha_{p,i}$ and $\alpha_{s,i}$ are given by
\begin{subequations}
	\label{eqn:condition_alpha}
	\begin{align}
		\label{eqn:condition_alphap}
		0 = f_{p,i}(\alpha) &\equiv \lsb \lp \omega_r-\omega_p + S_i -i\tfrac{\kappa}{2}\rp + \lp K + \tfrac{1}{3!} K_i \rp |\alpha_{p,i}|^2 + K'|\alpha_{p,i}|^4\rsb \alpha_{p,i} + \epsilon_p, \\
		\label{eqn:condition_alphas}
		0 = f_{s,i}(\alpha) &\equiv \lsb \lp \omega_r-\omega_s -i\tfrac{\kappa}{2}\rp + K |\alpha_{p,i}|^2 + K'|\alpha_{p,i}|^4\rsb \alpha_{s,i} + \epsilon_s.
	\end{align}
\end{subequations}
These choices ensure that the second term of $H_1'$ is zero and therefore eliminates the photon population in the polaron frame. We neglected frequency mixing terms and assumed that $|\alpha|^2\approx|\alpha_p|^2$ to obtain these conditions. We note that, from the point of view of a signal parametrically amplified by the resonator, \eqref{eqn:condition_alphap} contains the qubit signal $S_i+\tfrac{1}{3!} K_i|\alpha_{p,i}|^2$ directly as a parameter in the nonlinear equation for $\alpha_{p,i}$. This places this theory beyond linear response.

\subsubsection{Lamb shift} 
\label{ssub:lamb_shift}
When the conditions \eqref{eqn:condition_alpha} are met, the only remaining off-diagonal terms in $H''$ beside the spectroscopy drive is the Jaynes-Cummings coupling in $H_1''$. The only difference between $H''$ and the initial Hamiltonian ignoring the pump drive is that the qubit and resonator frequencies now depend on the amplitude of the cavity field $\alpha_p$, through the ac-Stark shift or the Kerr nonlinearity. Since this renormalized resonator is not driven, the photon population in this frame is zero and the Jaynes-Cummings coupling yields only a Lamb shift. Indeed, diagonalizing $H_1''$ using the usual dispersive transformation~\shortcite{Carbonaro1979} and projecting on the two-level $\{\ket{0},\ket{1}\}$ subspace yields
\begin{subequations}
	\begin{align}
		H''' &= \frac{\omega_1'''(\alpha)-\omega_0'''(\alpha)}{2}\sigma_z + g_0\lp \alpha_{s} e^{-i\omega_s t} \sigma_+ + \hc\rp + \omega_r'(\alpha)\ada, \\
		\dot\rho''' &= -i\comm{H'''}{\rho'''} + \gamma_\downarrow\sD[\sigma_-]\rho''' + \gamma_\uparrow\sD[\sigma_+]\rho''' + \frac{\gamma_\varphi'''}{2}\sD[\sigma_z]\rho''' + \kappa\sD[a]\rho''',
	\end{align}
\end{subequations}
where
\begin{subequations}
	\begin{align}
	\label{eqn:acstark}
		\omega_i'''(\alpha) &= \omega_i + S_i^p|\alpha_p|^2 + \frac{1}{4} K_i^p|\alpha_p|^4 + L_i^r(\alpha_p), \\
		L_i^r(\alpha) &= -g_{i-1}\lambda_{i-1}^r(\alpha), \\
		\lambda_i^r(\alpha) &= \frac{-g_i}{\omega_{i+1}''(\alpha)-\omega_{i}''(\alpha) - \omega_r'(\alpha)}, \\
		\gamma_{\downarrow} &= \gamma + \gamma_{\uparrow} + \lvb\lambda_0^r(\alpha_p)\rvb^2\kappa, \\
		\gamma_{\uparrow} &= \lp 2\gamma_\varphi + \kappa D^2 \rp |\lambda^p_0|^2 |\alpha_p|^2, \\
		\label{eqn:dephasingrate}
		\gamma_{\varphi}''' &= \gamma_\varphi + \frac{\kappa D^2}{2} + \frac{\gamma \lvb 2\chi_0^p\alpha_{p,0} - \chi_1^p\alpha_{p,1}\rvb^2}{2g_0^2}.
	\end{align}
\end{subequations}
$D=\lvb\alpha_{p,1}-\alpha_{p,0}\rvb$ is the distinguishability between the pointer states of the nonlinear resonator. The resulting master equation represents a qubit for which the linear $\propto S_i^p$ and quadratic $\propto K_i^p$ ac-Stark shifts depend on the pump frequency $\omega_p$ and the Lamb shift $L_i^r$ depends on the detuning between the Kerr-shifted resonator frequency $\omega_r'(\alpha)$ and the Stark-shifted qubit frequency. This qubit is subjected to modified relaxation rates.

The rate $\gamma_{\uparrow}$ contains dressed-dephasing as well as a very similar effect which we will refer to as dressed measurement-induced-dephasing. Dressed-dephasing represents population mixing caused by the qubit's pure dephasing bath when the qubit is dressed by photons~\shortcite{boissonneault_nonlinear_2008}. Dressed measurement-induced dephasing is the same effect when the dephasing occurs because of measurement-induced dephasing. As shown in~\shortcite{Boissonneault2009}, the precise value of these rates will depend on the baths' spectra. The third term of $\gamma_{\downarrow}$ is Purcell relaxation~\shortcite{Purcell1946,Houck2008}. In this nonlinear system however, the value of the Purcell rate depends on the Kerr-shifted resonator frequency and the Stark-shifted qubit frequency.

The second term of $\gamma_\varphi'''$ is measurement-induced dephasing~\shortcite{Schuster2005,Gambetta2006,Gambetta2008}. We recover the same expression as \shortciteNP{Gambetta2008}. Here however, the fields $\alpha_{p,i}$ are the solution of the nonlinear equation~\pref{eqn:condition_alphap}, in which the cavity-pull $S^p_i$ depends on qubit-pump detuning instead of the qubit-resonator detuning. Finally, the last term of $\gamma_\varphi'''$ is dressed-decay~\shortcite{boissonneault_nonlinear_2008}, and is negligible given the parameters of the experiment described in the next section.

\subsubsection{Discussion} 
\label{ssub:discussion}
Neglecting the last term in \eqref{eqn:dephasingrate} which vanishes in the limit of negligible spontaneous emission rate $\gamma$ yields $\gamma_{\varphi}'''= \gamma_\varphi + \frac{\kappa D^2}{2}$, the exact same equation as for a linear resonator [see \eqref{eq:linearbackaction}]. This remarkable non-trivial result indicates that the direct link between measurement-induced decoherence rate and distinguishability pertains even when the resonator is nonlinear, provided the approximations made in the previous paragraph are satisfied. Nevertheless the measurement-induced dephasing of a KNR is qualitatively different from the linear resonator case as we will see in the next section. The reason is that the pointer states amplitude, and therefore their distance in phase space, have a nonlinear dependence on the pump amplitude and frequency as expressed by \eqref{eqn:condition_alphap}.

In order to verify if the quantum limit \eqref{eq:ReadoutInequality} is reached here, $\Gamma_{\phi m}$ must be compared to the rate $\Gamma_{meas}$ at which information about the atom state is acquired. If the two pointer states are coherent states, then $\Gamma_{meas}=\kappa D^2$ as discussed in the beginning of this section. \eqref{eqn:dephasingrate} then implies that the backaction of the nonlinear resonator is at the quantum limit. According to the quantum theory of the pumped KNR however~\shortcite{Laflamme2011,drummond_quantum_1980}, this is true only in the limit of small parametric gain $G \simeq 1$, corresponding to $P_p$ being significantly different from $P_+ (\Omega)$. Indeed, when $G \gg 1$, the intracavity field shows enhanced phase-dependent fluctuations and the field reflected on the cavity shows some degree of squeezing. In that limit the measurement rate needs to be re-evaluated and our results do not allow us to conclude on the quantum-limited character of the pumped nonlinear resonator backaction.

In related work~\shortcite{Laflamme2011} Laflamme and Clerk make a different set of approximations that allows them to investigate the large gain limit. They assume that the qubit-resonator interaction can be described within linear response theory, which implies that $D$ is developed to first order in $\chi$ and that entanglement between the qubit and the resonator is neglected. In that framework, they are able to take into account the phase-dependent fluctuations of the resonator field and to treat the case of arbitrary parametric gain. With these approximations, they show that the measurement-induced dephasing rate is still given by \eqref{eqn:dephasingrate}. Moreover, they find that the measurement rate is $\Gamma_{meas} \simeq \kappa D^2 / G$ because of squeezing in the output field. The nonlinear resonator backaction on the qubit thus misses the quantum limit by a large factor of order $G$. 

We note however that typical experimental parameters in circuit QED are usually beyond the linear response theory due to the strong qubit-resonator coupling. Furthermore, the degree of validity of the linear response theory shrinks as the large-gain limit is approached, so that the conclusions in~\shortcite{Laflamme2011} are strictly speaking valid only in the limit where $\chi$ tends towards zero. To conclude, more theoretical work combining a polaron-type approach as in the previous section and an account of squeezing such as performed in~\shortcite{Laflamme2011} seems needed in order to assess rigorously the question of the quantum limit of the pumped KNR backaction on a strongly coupled qubit.


\subsection{Backaction of a nonlinear resonator on a qubit: measurements} 
\label{sub:measurements}

\begin{figure}[h]
\begin{center}
\hspace{0mm}
\includegraphics[width=12cm,angle=0]{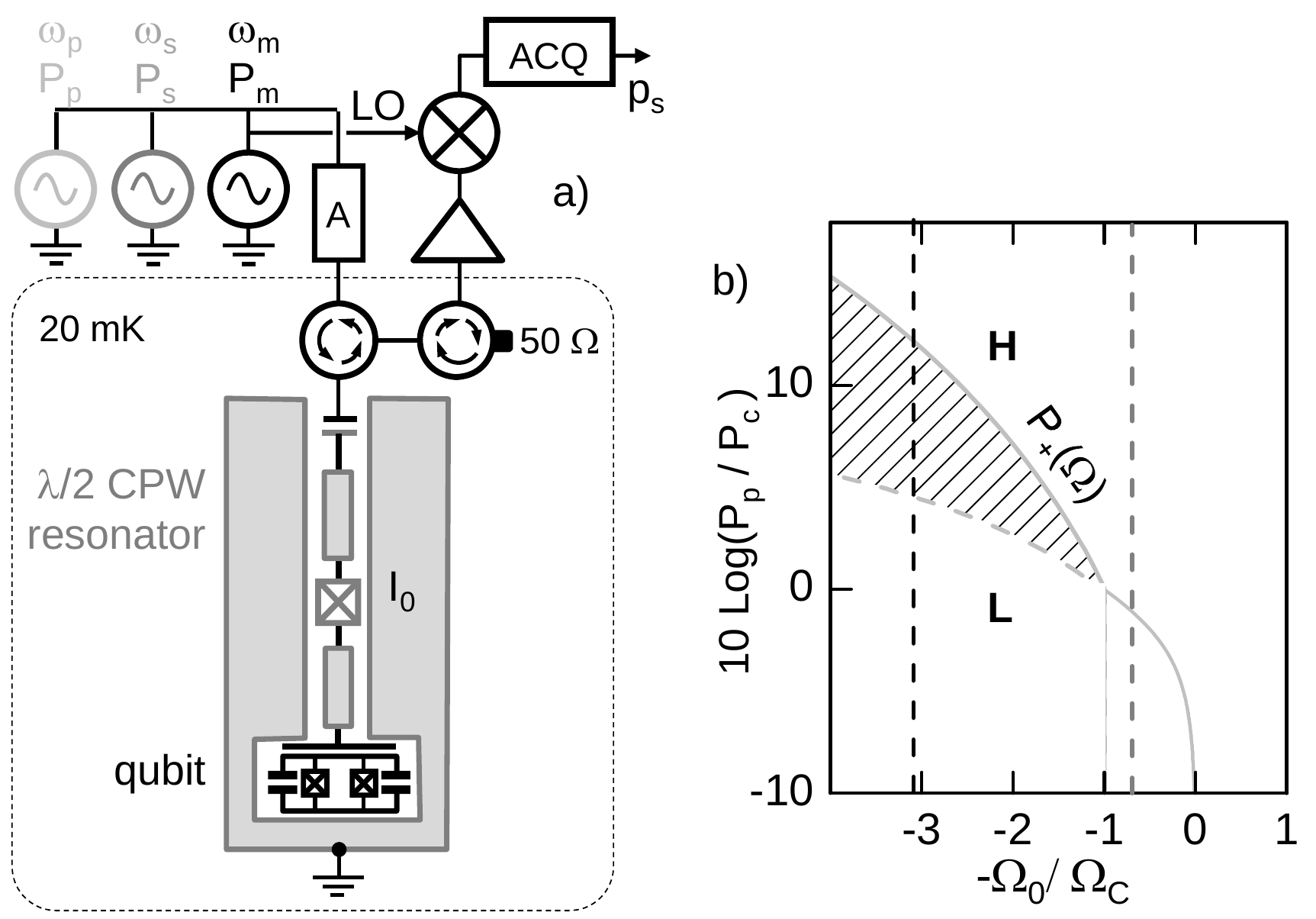}
\caption{(a)~Experimental setup: a transmon qubit is strongly coupled to a coplanar resonator made nonlinear with a Josephson junction. The sample is cooled to $20$~mK and is driven through an attenuator $A$ by three microwave sources. Source $p$ is used to establish a pump field in the resonator, source $s$ for qubit spectroscopy, and source $m$ as a JBA readout: its signal at $\omega_{m}$ is reflected from the resonator and routed through circulators to a cryogenic amplifier, a demodulator, and a digitizer, which yields the JBA switching probability $p_{s}$ and thus the probability of the qubit excited state $\ket{1}$. (b)~Stability diagram of the resonator in the  $\Omega$-$P_p$ plane. The solid lines indicate the highest PA gain below $\Omega_c$ and the power at which the resonator bifurcates from the low- (L) to the high- (H) amplitude state, respectively. The hatched area is the bistability region. Vertical dotted lines correspond the two datasets of Figs.~\ref{fig-acstark-fig2} and~\ref{fig-acstark-fig3}.}
\label{fig-acstark-fig1}
\end{center}
\end{figure}

We now present experimental data~\shortcite{Ong2011} aiming to test these predictions regarding the pumped nonlinear resonator backaction on a qubit that can be summarized as: 1) a qubit frequency shift given by Eqs.~\pref{eqn:acstark} and \pref{eqn:condition_alphap} 2) dephasing with a rate given by \eqref{eqn:dephasingrate}. For that purpose, we perform spectroscopy of the artificial atom in the presence of a pump field of variable power and frequency driving the KNR. The setup, illustrated in Fig.~\ref{fig-acstark-fig1}~a), is similar to the ones used in the previous section. In all that follows, the qubit-resonator detuning is fixed at $\Delta/2 \pi= 732$~MHz $\gg g/2\pi$, so that their interaction is well described by the dispersive Hamiltonian \eqref{eqn:dispersive_hamiltonian} with $\chi/2\pi=-0.8$~MHz and $\bar{s} / 2 \pi=1.7$~MHz. Qubit-state readout is performed as described in section~\ref{sec:cba_readout} at frequency $\omega_{m}/2\pi=6.439$~GHz ($\Omega_{0}/\Omega_c = 2$) by measuring the JBA switching probability $p_{\rm s}$.

\begin{figure}[t]
\begin{center}
\hspace{0mm}
\includegraphics[width=12cm,angle=0]{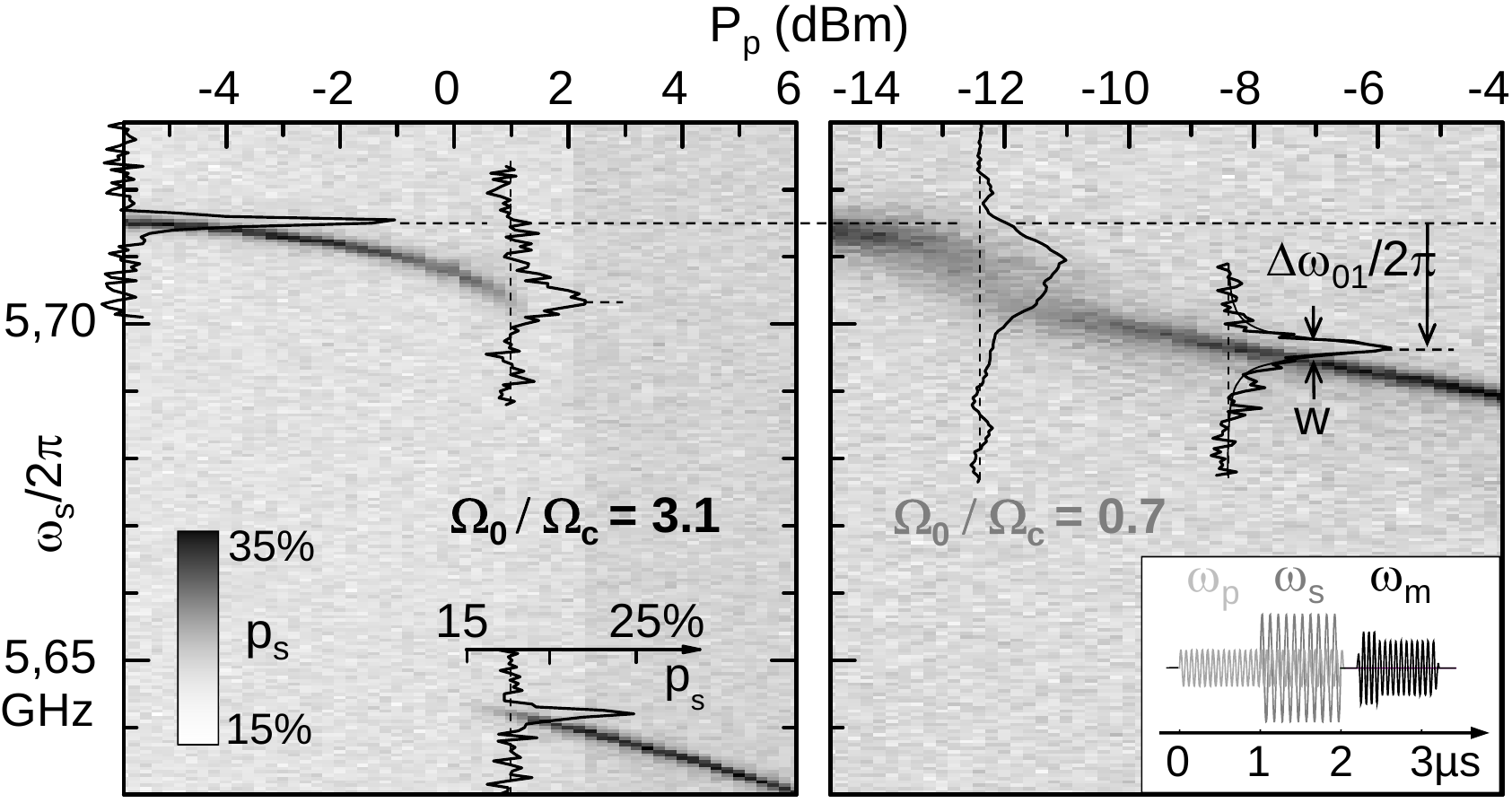}
\caption{2D plots of the switching probability $p_{\rm s}$ as a function of $\omega_{s}$ and $P_p$ for $\Omega_{0}/\Omega_c=3.1$ and $\Omega_0/\Omega_c=0.7$ (left and right panels). A few qubit lines are shown in overlay for negligible field amplitude in the resonator (top left), near the switching point at $P_p= 1.0$~dBm for $\Omega_0/\Omega_c=3.1$, and at  $P_p=-12.4$~dBm and $-8.6$~dBm for $\Omega_0/\Omega_c=0.7$. The horizontal dashed line indicates the qubit frequency at zero cavity field. Lorentzian fits of the qubit lines (see example at  $-8.6$~dBm) yield the ac-stark shift  $\Delta \omega_{01}$ and the FWHM linewidth $w$. Inset: microwave pulse sequence used.
}
\label{fig-acstark-fig2}
\end{center}
\end{figure}

The pulse sequence is shown in the inset of Fig.~\ref{fig-acstark-fig2}: we apply the spectroscopy pulse of varying frequency $\omega_{s}$ and fixed power $P_s$, while the resonator is driven by a microwave pump pulse of varying frequency $\omega_p$ and power $P_p$, followed at the end with a qubit readout pulse. Note that the pump pulse starts long before the spectroscopy pulse so that the intracavity field has reached its stationary state. Moreover, the readout pulse is applied $200$~ns after switching off both pump and spectroscopy pulses. This time is long enough to let the intraresonator field relax before readout, but shorter than the qubit relaxation time $T_{\rm 1}=700$~ns. Fig.~\ref{fig-acstark-fig2} shows the resulting qubit spectrum as a function of $P_p$ at two pump frequencies above and below $\Omega_c$, respectively in the parametric amplifier and the bistability regime. We observe the ac-Stark shift of the qubit resonance frequency towards lower values, as well as the broadening of the qubit line. This is akin to the behavior observed with a linear resonator, but with a very different dependence on $P_p$~\shortcite{Schuster2005}. First, an abrupt discontinuity in the ac-Stark shift at $\Omega_0/\Omega_c=3.1$ clearly indicates the sudden increase in the intracavity average photon number $\bar{n}$ as the resonator switches from $L$ to $H$. In this region, two spectroscopic peaks are observed at a given pump power. Second, the linewidth {\it narrows down} at large $\bar{n}$, in strong contrast with the linear resonator case where it increases linearly~\shortcite{Schuster2005}.

\begin{figure}[h]
\begin{center}
\hspace{0mm}
\includegraphics[width=6.5cm,angle=0]{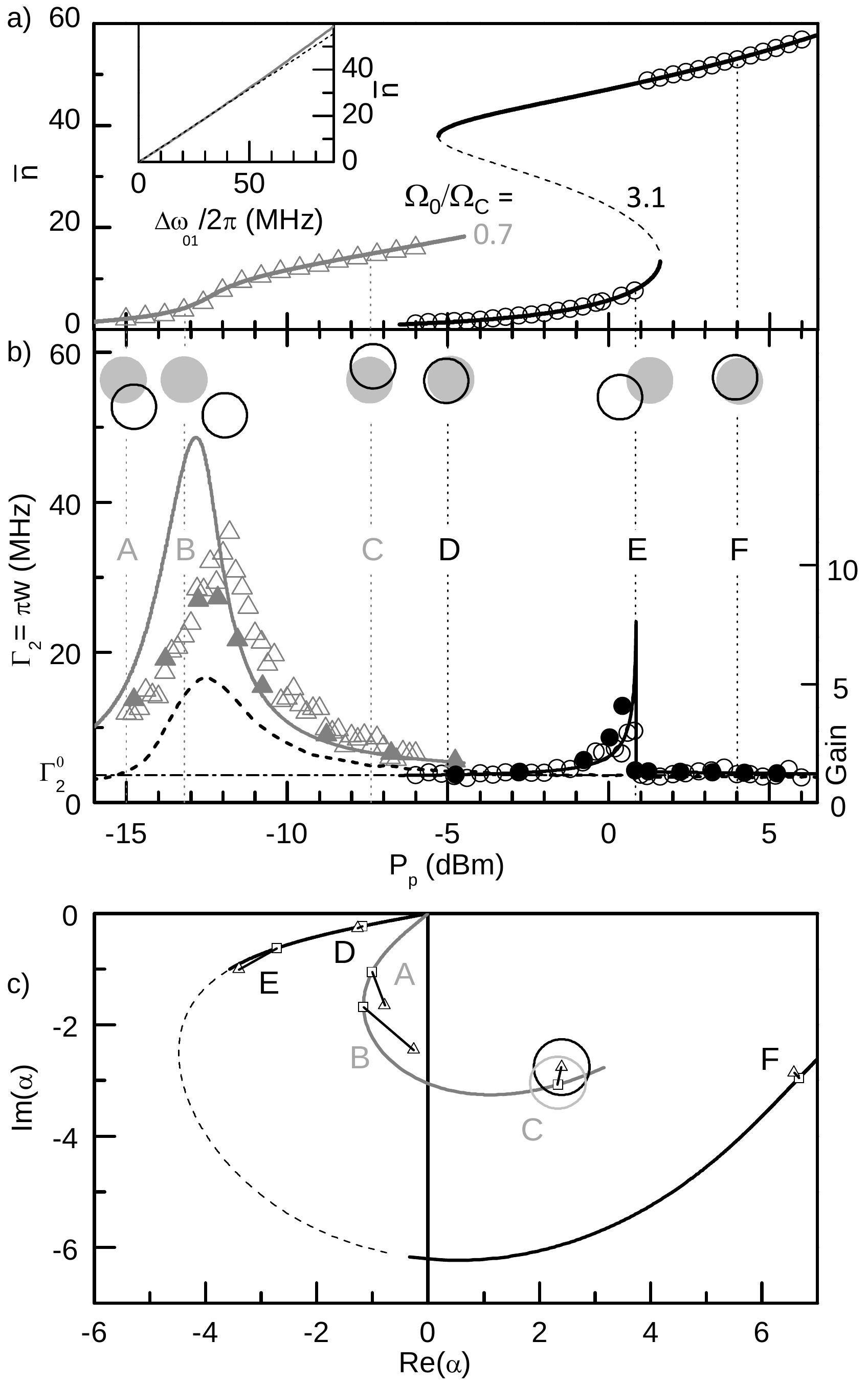}
\caption{(a) Experimental (open symbols) average photon number $\bar{n}$ and corresponding fits (lines - see text) as a function of $P_p$ for the same datasets  $\Omega_0/\Omega_c=3.1$ (circles) and $\Omega_0/\Omega_c=0.7$ (triangles) as in  Fig.~\ref{fig-acstark-fig2}. Experimental points are obtained by converting the measured $\Delta \omega_{01}$ according to the ac-Stark shift model of the inset (solid line - see text ; the dashed line shows the linear ac-Stark shift approximation). (b)~Qubit dephasing rate $\Gamma_2$  measured for the same datasets (open symbols) and calculated either by numerical integration of the system master equation (solid symbols) or using \eqref{eqn:dephasingrate} (solid lines). The horizontal dashed-dotted line indicates the intrinsic dephasing in zero cavity field. The measured parametric power gain  (dashed line) is also shown  for comparison for $\Omega_0/\Omega_c = 0.7$. (c)~Complex cavity field amplitude $\alpha_{\rm 0}$ (solid and dashed lines) calculated from \eqref{alpha} for increasing pump powers and for the same datasets. Both $\alpha_{\rm 0}$ (squares) and $\alpha_{\rm 1}$ (triangles) are also shown at six points labelled A-F. The separation between the two pointer states (segments) is to be compared to the uncertainty disk of a coherent state (open circles or disks shown at point C and at points A-F in panel b).}
\label{fig-acstark-fig3}
\end{center}
\end{figure}

Our measurements of the qubit frequency shift $\Delta \omega_{01} $ versus $P_p$ allow us to quantitatively test the theoretical predictions made in section \ref{sub:back-action_of_a_linear_resonator_on_a_qubit}. For that purpose, we use \eqref{eqn:acstark} to determine $\bar{n}=|\alpha_p|^2$ from $\Delta \omega_{01}$ using values for the $S_i^p$ and $K_i^p$ calculated for our qubit parameters [see Fig.~\ref{fig-acstark-fig3}~a) and inset]. We then fit the resulting experimental $\bar{n}(P_p)$ curves with the $\bar{n}$ values calculated from the sole dynamics of the resonator (with the qubit in $\ket{0}$) given by the square modulus of the solutions of \eqref{alpha} using $K$ and the measurement line attenuation $A$ as the only fitting parameters. The agreement is excellent over the whole $(\Omega,P_p)$ range for $K/2\pi=- 625\pm 15$~kHz and $A= 110.8\pm 0.2$~dB, which is consistent with the design value $K/2\pi = -750 \pm 250$~kHz and with independent measurements of the line attenuation $A=111\pm 2$~dB. This demonstrates that measuring the ac-Stark shift of a qubit is a sensitive method for probing the field inside a nonlinear resonator and for characterizing its Kerr nonlinearity~\shortcite{castellanos-beltran_widely_2007}.

Having obtained from the ac-Stark shift an accurate calibration of all the sample parameters including the microwave power reaching the sample, we are able to test quantitatively our predictions regarding the qubit dephasing rate induced by pumping the nonlinear resonator. We show in Fig.~\ref{fig-acstark-fig3}~b) the measured dephasing rate $\Gamma_2 = \pi w $ of the qubit, with $w$ the full width at half maximum (FWHM) of a Lorentzian fit to the experimental data. In addition to the measurement-induced dephasing $\Gamma_{\phi \rm m}$, $\Gamma_2$ also includes a constant contribution $\Gamma_2^0$ due to all other dephasing processes (mainly energy relaxation, flux noise and radiative broadening). For the sake of comparison, we also show the independently measured gain $G$ of the KR operated as a parametric amplifier on a small additional signal frequency-shifted by $100$~kHz from $\omega_p/2\pi$. Below and above $\Omega_c$, $\Gamma_2 (P_p)$ peaks where $G$ is maximum and near the bifurcation threshold respectively, and tends towards $\Gamma_2^0$ at large $\bar{n}$.

We also show the predictions of \eqref{eqn:dephasingrate} calculated without any adjustable parameter. The qualitative features of our data (the peak of the dephasing rate near the maximum of the gain or the bifurcation threshold) are well reproduced. The direct correlation between dephasing rate and distance in phase space between the nonlinear resonator pointer states (distinguishability $D$) is graphically evidenced in Fig.~\ref{fig-acstark-fig3}.

Despite this qualitative agreement, quantitative discrepancies are observed when $\Gamma_2$ is large. To understand their origin, we have calculated $\Gamma_2$ by numerical integration of the full master equation \eqref{eqn:system_masterequation} for the sample parameters and damping rates determined independently [see full symbols in Fig.~\ref{fig-acstark-fig3}~b)]. The agreement with the experimental data is excellent with no adjustable parameter over all parameters range, in contrast with experiments with a flux-qubit in which an unexplained shortening of the qubit relaxation time is observed when the KR is driven above bifurcation~\shortcite{Picot2008,Serban2010}. This implies that the discrepancies between our data and the predictions of \eqref{eqn:dephasingrate} are not due to an imprecision in determining our samples parameters, but only to the breakdown of the hypothesis $D\ll 1$ needed to derive this analytical formula, which is indeed only satisfied by our data when $\Gamma_2$ is below $\sim 10$~MHz. We thus confirm experimentally the theoretical predictions of section~\ref{sub:back-action_theory}: in the parameter range where the parametric gain is negligible ($G \simeq 1$) so that \eqref{eqn:acstark} is valid and the pointer states are coherent states, the backaction of the driven KNR on the qubit is close to the quantum limit. Further work is needed to study the large gain regime, and in particular to establish whether the qubit readout by a JBA discussed in section~\ref{sec:cba_readout} is quantum-limited or not. 



\section{Conclusion} 

In this chapter we have investigated experimentally and theoretically the coupling of a transmon qubit to a resonator with a Kerr nonlinearity, which constitutes one of the simplest extensions to the well-studied Jaynes-Cummings model. We have shown how effects arising from the resonator nonlinearity (bistability, parametric amplification, squeezing) qualitatively modify the dispersive qubit-resonator coupling. This allowed us in particular to obtain a qubit readout which has both a high fidelity and the potential of being QND. The quantum backaction exerted by the intraresonator field onto the qubit was also studied. We stress that despite the much more complex system dynamics, we obtain qualitative agreement with analytical calculations and quantitative agreement with numerical simulations, indicating that the sample parameters are controlled as well as in linear circuit QED. This opens the way to more elaborate experiments in which for instance the qubit could be used as a probe of the quantum fluctuations of the nonlinear resonator. In that way detailed tests of the theory of fluctuating nonlinear resonator could be performed.


\bibliographystyle{OUPnamed_notitle}
\bibliography{biblioQuantro110619}

\end{document}